\documentclass[12pt]{article}

\usepackage{a4wide}
\usepackage[dvips]{graphicx}

\newcommand{\nn}{\nonumber}
\newcommand{\be}{\begin{equation}}
\newcommand{\ee}{\end{equation}}
\newcommand{\ba}{\begin{eqnarray}}
\newcommand{\ea}{\end{eqnarray}}
\newcommand{\ci}[1]{\cite{#1}}
\def\vk{{\bf k}_{\perp}}

\def\vbs{{\bf b}}
\def\vb0{{\bf b}_0}
\newcommand{\LQCD}{\Lambda_{\rm{QCD}}}
\def\als{\alpha_s}

\def\mev{\,{\rm MeV}}
\def\gev{\,{\rm GeV}}

\def\xbj{x_{\rm Bj}}

\newcommand{\da}{{distribution amplitude}}
\newcommand{\wf}{wavefunction}
\newcommand{\lsim}{\raisebox{-4pt}{$\,\stackrel{\textstyle
                                                         <}{\sim}\,$}}
\newcommand{\gsim}{\raisebox{-4pt}{$\,\stackrel{\textstyle
                                                         >}{\sim}\,$}}

\newcommand{\tw}{\textwidth}
                                                         
\newcommand{\req}[1]{(\ref{#1})}
\def\xb{\bar{x}}

\def\={\,=\,}
\def\eps{\epsilon}
\def\veps{\varepsilon}

\begin{document}
\thispagestyle{empty}
\begin{flushright}
WU B 06-02 \\
hep-ph/0611290\\
November 2006\\[20mm]
\end{flushright}

\begin{center}
{\Large\bf The Longitudinal Cross Section of Vector Meson Electroproduction} \\
\vskip 15mm

S.V.\ Goloskokov
\footnote{Email:  goloskkv@theor.jinr.ru}
\\[1em]
{\small {\it Bogoliubov Laboratory of Theoretical Physics, Joint Institute
for Nuclear Research,\\ Dubna 141980, Moscow region, Russia}}\\
\vskip 5mm

P.\ Kroll \footnote{Email:  kroll@physik.uni-wuppertal.de}
\\[1em]
{\small {\it Fachbereich Physik, Universit\"at Wuppertal, D-42097 Wuppertal,
Germany}}\\

\end{center}

\vskip 15mm
\begin{abstract}
We analyze electroproduction of light vector mesons ($V=\rho, \phi$
and $\omega$) at small Bjorken-$x$ in the handbag approach in which the
process factorizes into general parton distributions and partonic 
subprocesses. The latter are calculated in the modified perturbative
approach where the transverse momenta of the quark and antiquark
forming the vector meson are retained and Sudakov suppressions are
taken into account. Modeling the generalized parton distributions 
through double distributions and using simple Gaussian wavefunctions 
for the vector mesons, we compute the longitudinal cross sections at 
large photon virtualities. The results are in fair agreement with the 
findings of recent experiments performed at HERA and HERMES.
\end{abstract}

\section{Introduction}
\label{sec:introduction}
Recently we analyzed light vector-meson electroproduction in the 
generalized Bjorken regime \ci{first}. That study bases on QCD
factorization \ci{rad96,col96} of the process $\gamma^*p\to Vp$ into
hard parton-level subprocesses  - meson electroproduction off partons
- and soft proton matrix elements representing generalized parton
distributions (GPDs). The subprocesses themselves factorize into hard
quark-antiquark pair production off partons - amenable to perturbation
theory - and soft $q\bar{q}$ transitions to the vector mesons. It
has been shown \ci{rad96,col96} that in this so-called handbag approach 
meson electroproduction is dominated by transitions from longitudinally
polarized virtual photons to vector mesons polarized alike. 
Other transitions are suppressed by inverse powers of the virtuality 
of the photon, $Q^2$. In Ref.\ \ci{first} we examined the kinematical
region accessible to the HERA experiments which is characterized by
high energies and very small values of Bjorken's variable, $\xbj$
($\lsim 10^{-2}$). In this region vector-meson electroproduction is
under control of the gluonic GPDs and the associated gluonic
subprocess $\gamma^*g\to Vg$; the quark GPDs play only a minor role.
In the present work we are going to extend our previous analysis 
\ci{first} to lower energies and to values of $\xbj$ up to about
0.2. This extension necessitates the inclusion of sea and valence
quark GPDs into the analysis as well as the associated subprocess
$\gamma^* q\to Vq$. As it turns out from our analysis, even in the
kinematical region accessible to the HERMES experiment, the gluonic 
GPD provides substantial contributions to vector-meson electroproduction.
This observation is in conflict with results from a previous attempt
\ci{goeke} where only the quark contributions have been
calculated within the handbag approach while the gluonic one has been
estimated from the leading-log approximation \ci{bro94} (where the
gluon GPD is approximated by the usual gluon distribution) and
added to the quark contribution incoherently. On the other hand, in
the recent leading-twist handbag analysis of meson electroproduction
performed by Diehl et al \ci{kugler} a relative strength of gluon and 
quark contributions has been found that is very similar to our
result. It is to be stressed however that the handbag approach
to leading-twist order grossly overestimates the longitudinal 
cross section in the kinematical region accessible to current
experiments. 

Here in this study we will restrict ourselves to the analysis of the 
longitudinal cross section, the most important and least
model-dependent observable of vector meson electroproduction. In
Sect.\ \ref{sec:handbag} we will briefly recapitulate the handbag
approach. In some detail we will only discuss the formation of
the vector meson from a quark-antiquark pair. In order to cure the
mentioned deficiencies of the leading-twist mechanism we will employ the
so-called modified perturbative approach \ci{li92} in which the quark
transverse degrees of freedom are retained and Sudakov suppressions
are taken into account. In Sect.\ \ref{sec:GPD} we construct the GPDs 
required in the factorization formula, from a reggeized ansatz for the 
double distribution \ci{rad98}. How we fix the parameters  that
specify the GPDs will be described in Sect.\ \ref{sec:parameters}
where also numerical results for the model GPDs are presented. The
comparison of our results for the longitudinal cross sections with 
experiment is left to Sect.\ \ref{sec:results}. Our summary is 
presented in Sect.\ \ref{sec:conclusions}. 
\section{The handbag amplitude}
\label{sec:handbag} 
\begin{figure}[t]
\begin{center}
\includegraphics[width=.35\tw,bb=113 221 299 340,%
clip=true]{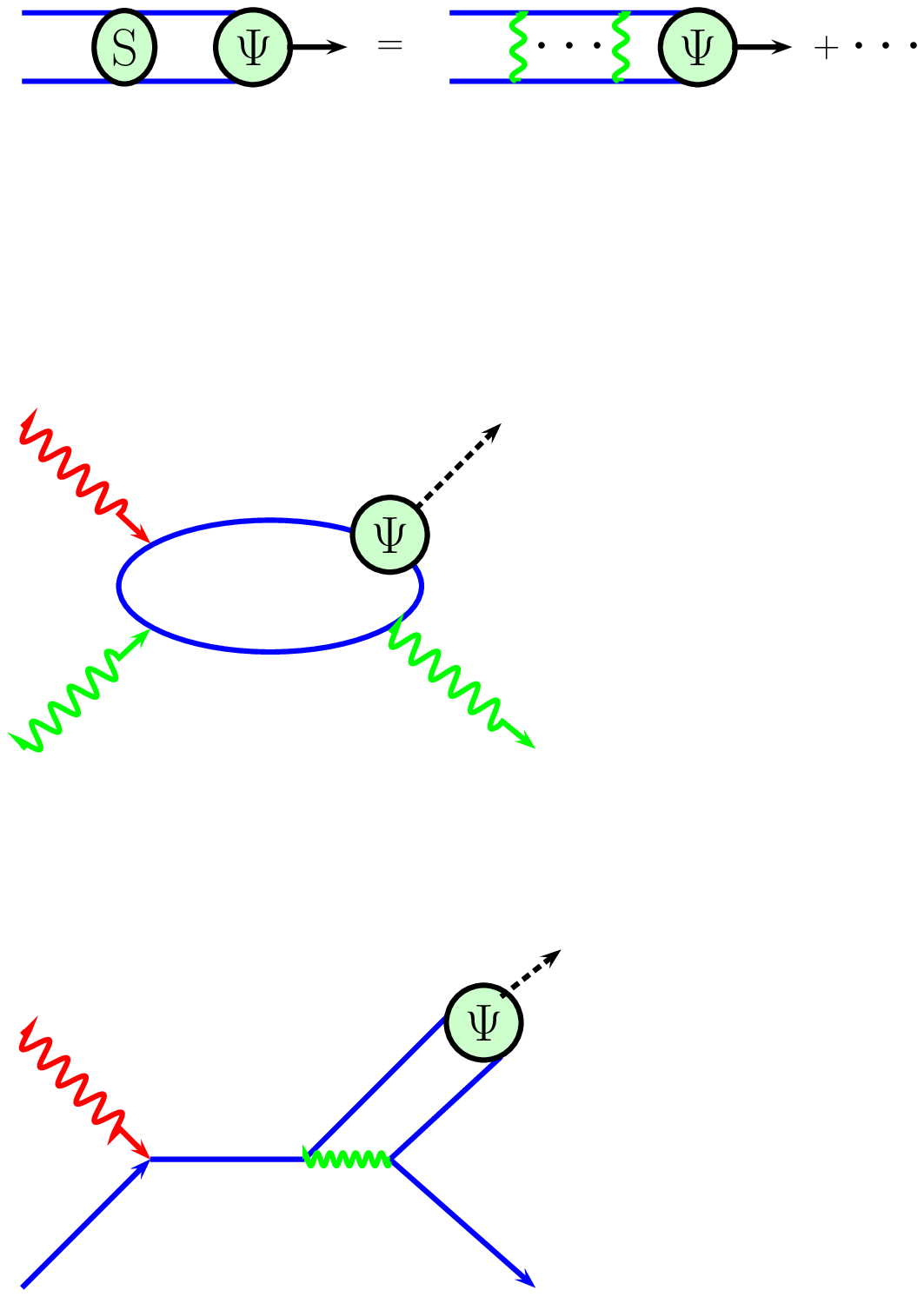} \hspace*{0.6cm}
\includegraphics[width=.35\textwidth,bb=113 391 312 530,%
clip=true]{meson-graph.ps}
\end{center}
\caption{Typical lowest order Feynman graphs for the two subprocesses
  of meson electroproduction.} 
\label{fig:feynman}
\end{figure}

We are interested in the process $\gamma^*p\to Vp$ for longitudinally
polarized photons ($\gamma^*_L$) and vector mesons ($V_L$). This
process can be extracted from electroproduction of vector mesons by
exploiting the familiar one-photon exchange approximation. We work in a
photon-proton center of mass system (c.m.s.), in a kinematical
situation where the c.m.s. energy, $W$, as well as the virtuality of the
photon are large while Bjorken's variable 
\be
\xbj\= \frac{Q^2}{W^2+Q^2-m^2}\,,
\label{xbj}
\ee 
is small ($\xbj\lsim 0.2$). The masses of the nucleon and the meson
are denoted by $m$ and $m_V$, respectively. Mandelstam $t$ is assumed
to be much smaller than $Q^2$. The proton has a rich structure. For
the dominant parton helicity non-flip configurations in the subprocess
there are four GPDs~\footnote{
Parton helicity flip configurations provide four more GPDs
\ci{diehl03}. Their neglect is vindicated by the properties of the
subprocess amplitudes which provide factors of either $-t/Q^2$ for the
gluonic subprocess or $\sqrt{-t}/Q$ for the quark one
\ci{first,hanwen}. The latter process is further suppressed by a twist-3
meson \wf{}.}
for each type of partons, named $H$, $\widetilde{H}$, $E$ and $
\widetilde{E}$. As discussed in detail in Ref.\ \ci{first} for 
unpolarized protons and small $\xbj$, respective small skewness~\footnote{
In Eq.\ \req{xi-xbj} also terms $\propto \xbj^2 m^2/Q^2$ and $\propto
\xbj t/Q^2$ occur which can safely be neglected in the kinematical
region of interest.}
\be
\xi \simeq \frac{\xbj}{2-\xbj}\,\big[1+m_V^2/Q^2\,\big]\,,
\label{xi-xbj}
\ee
only the GPD $H$ is to take into consideration, the three other ones 
do not contribute ($\widetilde{H}$, $\widetilde{E}$) or can be
neglected ($E$). Two subprocesses contribute to meson electroproduction, 
namely $\gamma^* g \to Vg$ and $\gamma^* q\to Vq$ (see Fig.\
\ref{fig:feynman}). This prompts us to decompose the 
$\gamma^*_L p \to V_Lp$ amplitude accordingly  
\be
{\cal M}_V \= {\cal M}_V^g + {\cal M}_V^q\,.
\label{amp}
\ee
The amplitude is normalized such that the partial cross section for
longitudinally polarized photons reads ($\Lambda$ is the usual
Mandelstam function)
\be
\frac{d\sigma_L}{dt} \= \frac1{16\pi (W^2-m^2) 
                   \sqrt{\Lambda(W^2,-Q^2,m^2)}}\,|{\cal M}_V|^2\,.
\label{sigma}
\ee
Strictly speaking this cross section also receives contributions from
the amplitudes for transitions from longitudinally polarized photons
to transversally polarized vector mesons. However, as the analysis of
the spin density matrix elements of the vector mesons reveal, see for
instance Refs.\ \ci{h1,zeus98,zeus05} or \ci{first}, this amplitude is 
very small and neglected by us. 

The gluonic contribution to the amplitude reads
\be
{\cal M}_V^g \= e \sum_a e_a{\cal C}_V^{\,a}\,\int_0^1\,
                       d\xb\,
                   {\cal H}_V^g(\xb,\xi,Q^2,t'=0)\,H^g(\xb,\xi,t')\,,
\label{gluon-amp}
\ee
while the quark one is
\be
{\cal M}_V^q\= e \sum_a e_a{\cal C}_V^{\,a}\, \int_{-1}^1\,
               d\xb\, {\cal H}_V^q(\xb,\xi,Q^2,t'=0)\,H^a(\xb,\xi,t')\,.
\label{quark-amp}
\ee
The sum runs over the quark flavors $a$ and $e_a$ denotes the quark
charges in units of the positron charge $e$. The non-zero flavor weight 
factors, ${\cal C}_V^a$, read 
\be
{\cal C}_\rho^{\,u}\=-{\cal C}_\rho^{\,d}\={\cal
  C}_\omega^{\,u}\={\cal C}_\omega^{\,d} 
       \=1/\sqrt{2}\,, \qquad {\cal C}_\phi^{\,s}\= 1\,.
\ee
Only the $t$ dependence of the GPDs is taken into account in the
amplitudes \req{gluon-amp} and \req{quark-amp}. That of the
subprocess amplitudes ${\cal H}$ provide corrections of order $t/Q^2$
which we neglect throughout this paper. In contrast to the subprocess
amplitudes the $t$ dependence of the GPDs is scaled by a soft parameter, 
actually by the slope of the diffraction peak. The full subprocess
amplitude, given for instance in Ref.\ \ci{hanwen}, indicate a breakdown of 
collinear factorization to order $t/Q^2$. As the argument of the GPDs 
we use $t'$ which is defined as
\be
t'\=t-t_{\rm min}\,,
\ee where
\be
t_{\rm min}\= -4m^2\,\frac{\xi^2}{1-\xi^2}\,,
\ee
is the minimal value of $t$ allowed in the process of interest. This
way we take into account a kinematical power correction. Since 
$t_{\rm min}$ is small, even at $\xi\simeq 0.1$ it only amounts to 
$-0.036\,\gev^{-2}$, this power correction, absorbed into the GPD, is
tiny. Also other power corrections of kinematical origin, as for
instance given in Eq.\ \req{xi-xbj} or in the phase space factor \req{sigma}, 
are taken into account by us. With the exception of these kinematical
effects hadron masses are otherwise neglected.

Let us now turn to the discussion of the subprocess amplitudes. As is
well-known, for the kinematics accessible to current experiments the
leading-twist contribution in which the partons are emitted and
reabsorbed by the protons collinearly and the meson is generated by 
one-gluon exchange in collinear approximation, does not suffice, 
see e.g.\ \ci{first,kugler}. The longitudinal cross section $\sigma_L$, 
i.e.\ the integrated differential cross section \req{sigma}, calculated 
to leading-twist order is well above experiment although with the
tendency of approaching experiment with increasing $Q^2$. In Sect.\ 
\ref{sec:results} we will return to this issue and present more
details on it. Another indication of the failure of the leading-twist 
order is the smallness of $R$, the famous ratio of the longitudinal 
and transversal cross sections. 

\begin{figure}[t]
\begin{center}
\includegraphics[width=.8\textwidth,bb=113 618 416 652,clip=true]
{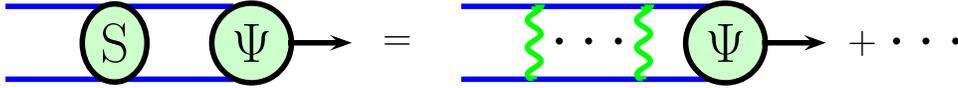}
\end{center}
\caption{Illustration of gluonic radiative corrections in axial gauge.} 
\label{fig:sudakov}
\end{figure}

As is well-known from extensive studies of electromagnetic form
factors at large momentum transfer leading-twist calculations are 
instable in the end-point regions since contributions from large 
transverse separations, $\vbs$, of quark and antiquark forming the 
meson are not sufficiently suppressed. In oder to eliminate that
defect the so-called modified perturbative approach has been invented
\ci{li92} in which the quark transverse degrees of freedom are
retained and the accompanying gluon radiation is taken into account. 
Thus, for the quarks and antiquarks entering the meson one allows 
for quark transverse momenta, $\vk$, with respect to the meson's 
momentum. In addition, as suggested in  Ref.\ \ci{jak93}, one also 
makes allowance for a meson light-cone \wf{} $\Psi_V(\tau, k_\perp)$ 
where $\tau$ is the fraction of the light-cone plus component of the 
meson's momentum the quark carries; the antiquark carries the fraction
$\bar{\tau}\=1-\tau$. In Ref.\ \ci{li92} the gluon radiation has been 
calculated in the next-to-leading-log (NLL) approximation using 
resummation techniques and having recourse to the renormalization 
group. The quark-antiquark separation $\vbs$ in configuration space 
acts as an infrared cut-off parameter. Radiative gluons with wave
lengths between the infrared cut-off and an upper limit (related to 
the hard scale $Q^2$) yield suppression, softer gluons are part of the
meson \wf{} while harder ones are an explicit part of the subprocess 
amplitude. Congruously, the factorization scale is given by the 
quark-antiquark separation, $\mu_F=1/b$, in the modified perturbative 
approach. In axial gauge the Sudakov factor can be regarded as a 
modification of the meson's \wf{} \ci{li92} in a fashion that is 
depicted in Fig.\ \ref{fig:sudakov}.

Here in this work we are going to employ the modified perturbative
approach too. In contrast to Ref.\ \ci{goeke} we still consider the
partons entering the subprocess as being emitted and reabsorbed
by the proton collinearly. This supposition relies on the fact that
all Fock states of the proton contribute to the GPDs. Hence, the
r.m.s.\ $\vk$ of the partons inside the proton reflects the charge 
radius of the proton (i.e.\ $\langle \vk^2\rangle^{1/2} \simeq
200\,\mev$). Only a mild $\vk$ dependence of the GPDs is therefore to
be expected. This is to be contrasted with the situation for the meson
where the hard process only feeds its valence Fock state. The
compactness of the latter entails much larger values of $\vk$.
The modified perturbative approach applied to the subprocess, is to
some extent similar to the mechanism proposed in Ref.\ \ci{fra95} for 
the suppression of the gluon contribution to meson electroproduction
in the leading $\ln(1/\xbj)$ approximation \ci{bro94}.

Since the resummation of the logs involved in the Sudakov factor can
only efficiently be performed in the impact parameter space \ci{li92}
we have to Fourier transform the lowest-order subprocess amplitudes
to that space and to multiply them there with the Sudakov factor. This
leads to ($i=g,q$)
\be
{\cal H}_V^i \= \int d\tau d^2b\, 
         \hat{\Psi}_V(\tau,-\vbs)\, \hat{\cal F}_V^i(\xb,\xi,\tau, Q^2,\vbs)\, 
         \als(\mu_R)\,{\rm exp}{[-S(\tau,\vbs,Q^2)]}\,.
\label{mod-amp}
\ee
The two-dimensional Fourier transformation between the canonical conjugated
$\vbs$ and $\vk$ spaces is defined by 
\be
    \hat{f}(\vbs) \= \frac{1}{(2\pi)^2} \int d^{\,2}\, \vk\, \exp{[-i
                                  \,\vk\cdot\vbs\,]}\; f(\vk)\,.
\label{fourier}
\ee
The Sudakov exponent $S$ in (\ref{mod-amp}) is given by \ci{li92}
\be
S(\tau,b,Q)=s(\tau,b,Q)+s(\bar{\tau},b,Q)-\frac{4}{\beta_0}
\ln\frac{\ln\left(\mu_R/\Lambda_{QCD}\right)}
                                {\hat{b}}\,,
\label{sudeq}
\ee
where a Sudakov function $s$ occurs for each quark line entering the
meson and the abbreviation 
\be
  \hat{b} \= -\ln{\left(b\LQCD\right)} \,,
\ee
is used. The last term in (\ref{sudeq}) arises from the application of 
the renormalization group equation ($\beta_0=11-\frac{2}{3} n_f$) where
$n_f$ is the number of active flavors. A value of $220\,\mev$ for 
$\Lambda_{QCD}$ is used here and in the evaluation of $\als$ from the 
one-loop expression. The renormalization scale $\mu_R$ is taken to be 
the largest mass scale appearing in the hard scattering amplitude,
i.e. $\mu_R=\max\left(\tau Q, \bar{\tau} Q,1/b\right)$. This choice
avoids large logs from higher orders pQCD. Since the 
bulk of the handbag contribution to the amplitudes is accumulated in
regions where $\mu_R$ is smaller than $3\,\gev$ we have to deal with
three active flavors, i.e.\ we take $n_f=3$. For small $b$ there is no 
suppression from the Sudakov factor; as $b$ increases the Sudakov
factor decreases, reaching zero at $b=1/\Lambda_{QCD}$. For even
larger $b$ the Sudakov is set to zero~\footnote
{The definition of the Sudakov factor is completed by the following
rules \cite{li92}: $\exp{[-S]}=1$ if $\exp{[-S]}\geq 1$,
$\exp{[-S]}=0$ if $b\geq 1/\LQCD$ and $s(\beta,b,Q)=0$ if $b\leq
\sqrt{2}/\beta Q$.}. 
The Sudakov function $s$ reads
\be
 s(\tau,b,Q) = \frac{8}{3\beta_0} \left( \hat{q} 
                       \ln{\left(\frac{\hat{q}}{\hat{b}}\right)}
                       -\hat{q} + \hat{b} \right) + {\rm NLL-terms}\,,
\label{eq:sudakov}
\ee
where
\be
     \hat{q} \= \ln{\left(\tau Q/(\sqrt{2}\LQCD)\right)}\,.
\ee
Actually we do not use the version of the NLL terms quoted in Ref.\ 
\ci{li92} but rather that one given in Ref.\ \ci{DaJaKro:95}. The 
latter one includes some minor corrections which are hardly relevant 
numerically. Due to the properties of the Sudakov factor any 
contribution to the amplitudes is damped asymptotically, i.e. for 
$\ln (Q^2/\LQCD^2)\to\infty$, except those from configurations with 
small quark-antiquark separations. 

The hard scattering kernels ${\cal F}_V^i$ or their Fourier
transform $\hat{{\cal F}}_V^i$ occurring in Eq.\ \req{mod-amp} are
computed from the pertinent Feynman graphs, see Fig.\ \ref{fig:feynman}. 
The result for the gluon subprocess is discussed in some detail in
Ref.\ \ci{first} and we refrain from repeating the lengthy expressions
here. For quarks, on the other hand, the hard scattering kernel reads
\be
{\cal F}_V^q \= C_F \sqrt{\frac{2}{N_c}}\,\frac{Q}{\xi}\,
           \left[\frac1{k_\perp^2 + \bar{\tau}(\xb+\xi)Q^2/(2\xi)-i\eps}
          -\frac1{k_\perp^2 -\tau(\xb-\xi)Q^2/(2\xi) - i\eps}\right]\,,
\label{q-kernel}
\ee
where $N_c$ denotes the number of colors and $C_F=(N_c^2-1)/(2N_c)$ is
the usual color factor. Here and in the gluon kernel as well we only 
retain $k_\perp$ in the denominators of the parton propagators 
where it plays a crucial role. Its square competes with terms 
$\propto \tau (\bar{\tau}) Q^2$ which become small in the end-point
regions where either $\tau$ or $\bar{\tau}$ tends to zero.

The denominators of the parton propagators in \req{q-kernel} and in
the gluonic kernel are either of the type
\be
   {T}_1 \= \frac{1}{\vk^2 + d_1Q^2}\,,
\ee
or
\be
T_2 \= \frac1{\vk^2 - d_2 (\xb\pm\xi) Q^2 - i \veps}\,.
\label{eq:T2}
\ee
where $d_i\geq 0$. The Fourier transforms of these propagator terms can 
readily be obtained:
\ba
\hat{T}_1 &=& \frac1{2\pi}\, K_0(\sqrt{d_1}\,bQ)\,,\nn\\
\hat{T}_2 &=& \frac{1}{2\pi}\, 
           K_0\left(\sqrt{d_2(\pm\xi -\xb)}\, bQ\right)\;
                    \theta(\pm\xi -\xb) \nn\\
             &+&  \frac{i}{4} \, 
             H^{(1)}_0\left(\,\sqrt{d_2\,(\,\xb\pm\xi\,)}\,bQ\right)\; 
                        \theta(\,\xb\pm\xi\,)\,,
\label{eq:T2-FT}
\ea
where $K_0$ and $H_0^{(1)}$ are the zeroth order modified Bessel
function of second kind and Hankel function, respectively. 

For the purpose of comparison we also quote the leading-twist results
(i.e.\ the limit $k_\perp\to 0$) for the subprocess amplitudes
\ba
{\cal H}_{V({\rm l.t.})}^g &=& \frac{8\pi\als}{N_c Q} f_V 
  \langle 1/\tau\rangle_V\, \Big[(\xb-\xi)(\xb-\xi+i\veps)\Big]^{-1}\,, \nn\\
{\cal H}_{V({\rm l.t.})}^q &=& \frac{4\pi\als}{N_c Q} f_V 
    \langle 1/\tau\rangle_V\,C_F\,\Big[\frac1{\xb-\xi+i\eps} +
      \frac1{\xb+\xi-i\eps}\Big]\,.
\label{lt}
\ea 
Here the $f_V$ denote the decay constants of the vector mesons for
which we adopt the values $f_\rho \= 209\,\mev$, $f_\phi\=221\,\mev$
and $f_\omega\=187\,\mev$ from Ref.\ \ci{neubert}. The $1/\tau$ moment 
of the meson's distribution amplitude which represents its \wf{} 
integrated over $k_\perp$ (up to the factorization scale), is denoted 
by $\langle 1/\tau\rangle_V$. In deriving Eq.\ \req{lt} use is made of
the fact that the distribution amplitudes for the vector mesons we are 
interested in are symmetric under the exchange 
$\tau \leftrightarrow \bar{\tau}$.

\section{Modeling the GPDs}
\label{sec:GPD}
As in Ref.\ \ci{first} we construct the GPDs with the help of the
double distributions invented by Radyushkin \ci{rad98}. The main
advantage of this construction is the warranted polynomiality of the
resulting GPDs and the correct forward limit $\xi, t \to0$. It is
well-known that at low $x$ the  parton distribution functions (PDFs) 
behave as powers $\delta_i$ of $x$. These powers are assumed to be 
generated by Regge poles \ci{feynman} and for sea and valence
quarks they are identified with the ususal Regge intercepts 
$\alpha_i(0)$. As a consequence of the familiar definition of the 
gluon GPD which reduces to $xg(x)$ in the forward limit, the power 
$\delta_g$ is shifted by -1 as againts the intercept of the 
corresponding Regge trajectory. We generalize this behavior of the 
PDFs by assuming that the $t'$ dependence of the GPDs is also under
control of Regge behavior (for a similar ansatz see Ref.\ \ci{guzey}). 
In order to simplify matters we consider 
only two trajectories: a gluon one, $\alpha_g(t')$, which is
understood as the hard physics partner of the famous Pomeron
trajectory playing a prominent role in soft high energy diffractive 
scattering, and a Regge trajectory for the valence  quarks, 
$\alpha_{\rm val}(t')$. This trajectory represents the family of the 
leading Regge poles that couple to the valence quarks of the proton 
($\rho$, $\omega$, $a_2$ and $f_2$ exchange in soft scattering). 
Since the sea quarks mix with the gluons under evolution it seems 
plausible to use the gluon trajectory for the sea quark GPDs 
as well. This assumption is supported to some extent by the
observation that the ratios of the various sea quark PDFs over the
gluon one are approximately independent of $Q^2$ at low $x$ and for 
$Q^2 \gsim 4\,\gev^2$. The trajectories are assumed to be linear 
functions of $t'$ in the small-$t'$ range
\be
 \alpha_i \= \alpha_i(0) + \alpha'_i t'\,, \qquad \qquad i=g, {\rm val}\,,
\ee 
and are accompanied by Regge residues assumed to have an exponential
$t'$ dependence with slope parameters $b_i$.

Following Radyushkin \ci{rad98} but generalizing to non-zero $t'$, we
employ the following ansatz for the double distributions ($i=g$, sea,
val)
\be
f_i(\beta,\alpha,t')\= {\rm e}^{(b_i+\alpha_i'\ln(1/|\beta|))t'}\, h_i(\beta)\,
                   \frac{\Gamma(2n_i+2)}{2^{2n_i+1}\,\Gamma^2(n_i+1)}
                   \,\frac{[(1-|\beta|)^2-\alpha^2]^{n_i}}
                           {(1-|\beta|)^{2n_i+1}}\,,
\label{DD}
\ee
where
\ba
h_g(\beta)\; &=& |\beta|g(|\beta|)\, \hspace*{0.154\textwidth} n_g\;=2\,, \nn\\
h^q_{\rm sea}(\beta) &=& q_{\rm sea}(|\beta|)\;{\rm sign}(\beta)
                         \hspace*{0.07\textwidth} n_{\rm sea}=2\,, \nn\\
h^q_{\rm val}(\beta) &=& q_{\rm val}(\beta)\, \Theta(\beta)    
                   \hspace*{0.117\textwidth} n_{\rm val}=1\,.
\label{function-h}
\ea
For the decomposition of the double distribution into valence and sea
contribution we follow the procedure proposed in Ref.\ \ci{diehl03}
(for earlier discussions on this decomposition see \ci{goeke,bel02}) and
write 
\ba
f^q_{\rm val}(\beta,\alpha,t') &=&\big[f^q(\beta,\alpha,t')
              + f^q(-\beta,\alpha,t')\big]\, \Theta(\beta)\,, \nn\\
f^q_{\rm sea}(\beta,\alpha,t') &=&f^q(\beta,\alpha,t')\,\Theta(\beta)\,
                -\,f^q(-\beta,\alpha,t')\,\Theta(-\beta)\,.
\ea
In the forward limit ($\xi, t'\to 0$) this decomposition is conform to
the usual definition of sea and valence quark PDFs.

According to Ref.\ \ci{rad98} the GPDs are related to the double
distributions by the integral
\be
H_i(\xb,\xi,t')\=\int_{-1}^1 d\beta\,\int_{-1+|\beta|}^{1-|\beta|} d\alpha\,
                 \delta(\beta+\xi\alpha-\xb)\,f_i(\beta,\alpha,t')\,.
\label{GPD-DD}
\ee
For convenience we employ an expansion of the PDFs~\footnote{
These terms may be interpreted as a set of daughter trajectories
spaced by $1/2$.}
\be
h_i(\beta) \=
\beta^{-\delta_i}\,(1-\beta)^{\,2n_i+1}\;\sum_{j=0}^3\,c_{ij}\,\beta^{j/2}\,,
\label{pdf-exp}
\ee
which is particularly useful at low $\beta$ and allows to perform the
integral \req{GPD-DD} term by term analytically. This results in a
corresponding expansion of the GPDs 
\be
H_i(\xb,\xi,t') \= {\rm e}^{b_it'}\,\sum_{j=0}^3\, c_{ij}\,H_{ij}(\xb,\xi,t')\,.
\label{GPD-exp}
\ee
For gluons and sea quarks~\footnote{
    Because $\delta_{\rm sea}>1$ there
    is a singularity at $\beta=0$ in the integral \req{GPD-DD}. As
    suggested by Radyushkin \ci{rad98} this singularity is regularized by
    considering \req{GPD-DD} for $H^q_{\rm sea}$ as a principal value integral.} 
the individual terms read ($\eps_g=1$, $\eps_{\rm sea}=-1$,
$m_{gj}=3+j/2-\delta_g -\alpha_g't$, $m_{{\rm sea}\,j}=m_{gj}-1$) \ci{rad98} 
\ba
H_{ij}(\xb,\xi,t') &=& \frac{15}{2\xi^5}\, 
                \frac{\Gamma(m_{ij}-2)}{\Gamma(m_{ij}+3)}\,
                \left\{\left[(m_{ij}^2+2)(\xi^2-\xb)^2 -
                    (m_{ij}^2-1)(1-\xi^2)(\xb^2-\xi^2)\right] \right. \nn\\
              &\times& \left.\left[x_1^{m_{ij}}- x_2^{m_{ij}}\right]
                + 3m_{ij}\, \xi(1-\xb)(\xi^2-\xb)  
                  \left[x_1^{m_{ij}}+ x_2^{m_{ij}}\right]\right\} \qquad
                  \xb\geq \xi \nn\\
            &=& \frac{15}{2\xi^5}\, \frac{\Gamma(m_{ij}-2)}{\Gamma(m_{ij}+3)}\,
                \left\{ x_1^{m_{ij}}\left[(m_{ij}^2+2)(\xi^2-\xb)^2 +
                 3m_{ij}\,\xi(\xi^2-\xb)(1-\xb) \right. \right. \nn\\
            &&\left.\left. -(m_{ij}^2-1)(1-\xi^2)(\xb^2-\xi^2)\right]
              + \eps_i (\xb \to -\xb)\right\} \qquad 0 \leq \xb <
            \xi\,,
\label{model2}
\ea
and for valence quarks 
($m_{{\rm val}\,j}=2+j/2-\delta_{\rm val}-\alpha'_{\rm val}t$)
\ba
H_{{\rm val}\,j}(\xb,\xi,t') &=& \frac{3}{2\xi^3}\,
\frac{\Gamma(m_{{\rm val}\,j}-1)}{\Gamma(m_{{\rm val}\,j}+2)}\, \nn\\
             &\times&  \left\{(\xi^2-\xb)(x_1^{m_{{\rm val}\,j}} -
                 x_2^{m_{{\rm val}\,j}})
                  + m_{{\rm val}\,j}\,\xi(1-\xb) (x_1^{m_{{\rm val}\,j}} 
               +x_2^{m_{{\rm val}\,j}})\right\} \quad\, \xb\geq \xi\nn\\
           &=& \frac{3}{2\xi^3}\,\frac{\Gamma(m_{{\rm
               val}\,j}-1)}{\Gamma(m_{{\rm val}\,j}+2)}\, 
        x_1^{m_{{\rm val}\,j}} \left[\xi^2-\xb + m_{{\rm val}\,j}\,\xi(1-\xb)\right]
                \quad -\xi\leq \xb <\xi\,.
\label{model1}
\ea
In Eqs.\ \req{model1} and \req{model2} we use the short-hand expressions
\be
x_1\=\frac{\xb+\xi}{1+\xi}\,, \qquad x_2\=\frac{\xb-\xi}{1-\xi}\,, 
\qquad x_3\=\frac{\xb-\xi}{1+\xi}\,.
\ee
The definition of the GPDs is completed by the relations
\be
H^g(-\xb,\xi,t') \= H^g(\xb,\xi,t')\,, \qquad H_{\rm
  sea}^q(-\xb,\xi,t')\=-H_{\rm sea}^q(\xb,\xi,t')\,,
\label{symmetry}
\ee
and
\be
H_{\rm val}^q(\xb,\xi,t') \=0\, \hspace*{0.1\textwidth} -1 \geq \xb <
-\xi\,.
\ee
The GPDs and their derivatives up to order $n_i$ are continuous at 
$\xb\=\xi$. In the present work we do not consider the full evolution 
of the GPDs. We rather approximate it by that of the PDFs. Since for
$\xi\ll\xb$ and $t'\simeq 0$ the GPDs turn into the corresponding PDFs
up to corrections of order $\xi^2$, evolution is therefore taken into
account in that region approximately. The corrections to the PDFs are of different
size - they are very small for the gluon and largest for the sea quarks.
The approximative treatment of evolution receives further support by
the fact that at low skewness the dominant contribution to the
longitudinal cross section is provided by the imaginary part of the
amplitude \req{amp}. The imaginary part is related to 
the GPDs at $\xb\simeq\xi$ (here differences between $\xb$ and $\xi$ are of 
order $\langle k^2_\perp\rangle/Q^2$) which read for $\xi\ll 1$:
\ba
H^g(\xi,\xi,t') &=& \frac{  2\xi g(2\xi)\;
   {\rm e}^{(b_g+\alpha_g'\ln{[(1+\xi)/(2\xi)]})t'}}   
                    {(1-\delta_g/5-\alpha_g't'/5)(1-\delta_g/4-\alpha_g't'/4)
                     (1-\delta_g/3-\alpha_g't'/3)}\,, \nn\\[1em]
H^q_{\rm val}(\xi,\xi,t') &=& \frac{q_{\rm val}(2\xi)\;  {\rm e}^{(b_{{\rm
         val}}+\alpha_{\rm val}'\ln{[(1+\xi)/(2\xi)]})t'}} 
                {(1-\delta_{\rm val}/3-\alpha_{\rm val}'t'/3)
                 (1-\delta_{\rm val}/2-\alpha_{\rm val}'t'/2)}\,.
\label{xixi}
\ea
For sea quarks one has to replace $\delta_g$ by $\delta_{\rm sea}=1+\delta_g$ 
in the first equation. Hence, at least at small $\xi$, evolution of
the GPDs is approximately taken into account. The implementation of
the full evolution is left to a forthcoming paper. From Eq.\ \req{xixi} 
one may also read off the so-called skewing effect \ci{bel02,mrt,man98}, 
i.e.\ the enhancement of the GPDs at $\xb=\xi$ and at $t'=0$ over the 
corresponding PDFs taken at the momentum fraction $2\xi$. For the
gluons the skewing effect provides the difference between the
leading-twist result for vector meson electroproduction and the 
leading-log approximation \ci{bro94} for which $H^g(\xi,\xi,t'\simeq 0)$ 
is replaced by $2\xi g(2\xi)$. It is easy to see from Eqs.\
\req{GPD-exp}, \req{model2} and \req{xixi} that the leading-log 
approximation is only valid at very low $\xi$. One may also see from 
Eq.\ \req{xixi} that the use of the $n=1$ model for the sea quarks 
would lead to implausibly large values for $H_{\rm sea}(\xi,\xi,t')$. 
This is another reason why we prefer the $n=2$ GPD model for the sea
quarks.
 
In Eq.\ \req{GPD-DD} so-called $D$ terms for gluons and sea quarks are 
ignored \ci{pol99}. The $D$ terms ensure the appearance of the highest
powers of the skewness in the moments of the GPDs. According to the 
above discussion the $D$ terms only contribute to the less important 
real part of the amplitude since their support is the region 
$-\xi \le \xb \le \xi$. We take this in vindication of neglecting 
the $D$ terms.

\section{Fixing the parameters}
\label{sec:parameters}
\begin{figure}[t]
\begin{center}
\includegraphics[width=.40\textwidth,bb=139 362 461 701,%
clip=true]{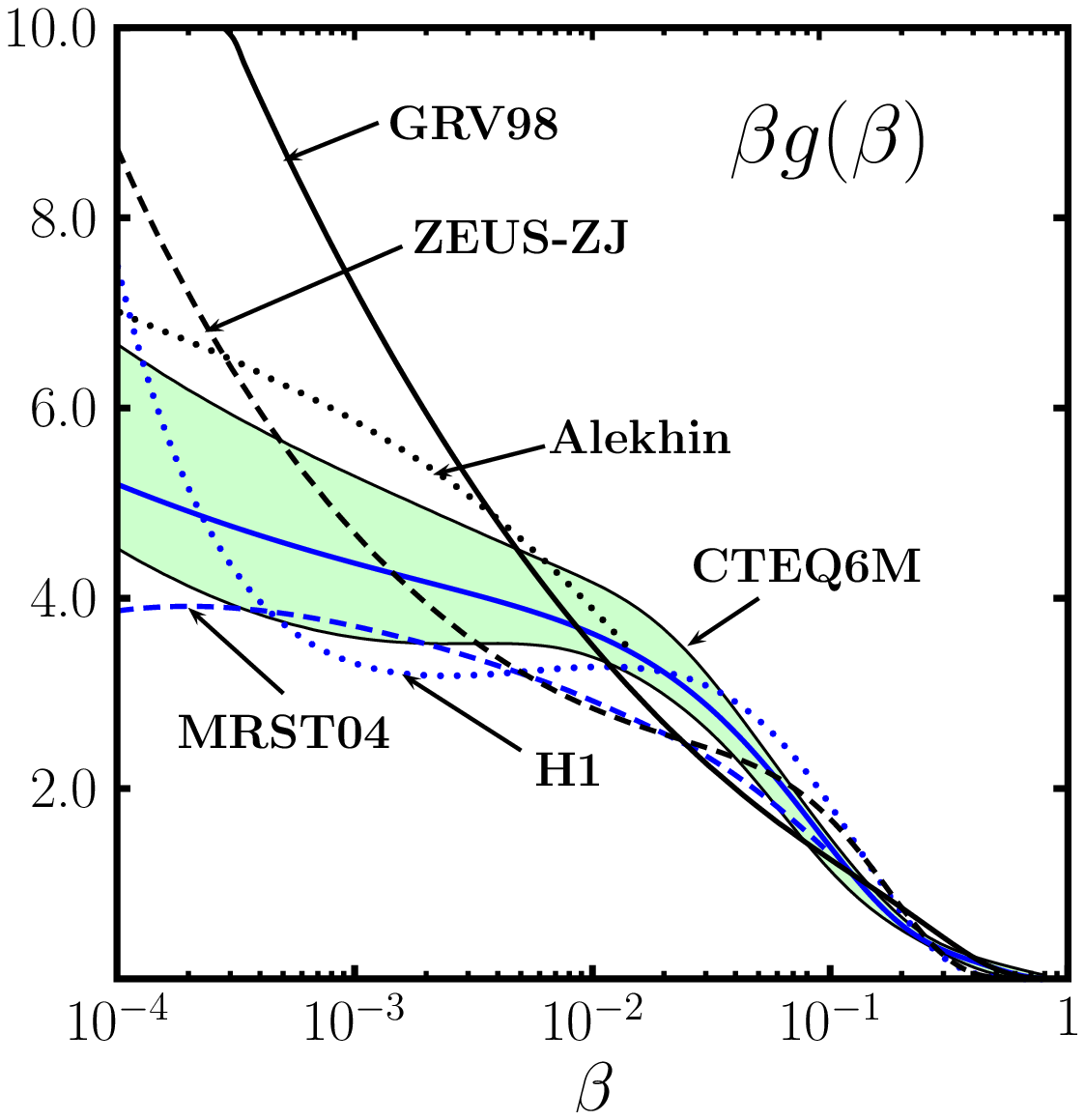} \hspace*{1.0em}
\includegraphics[width=.41\textwidth,bb=98 316 430 658,%
clip=true]{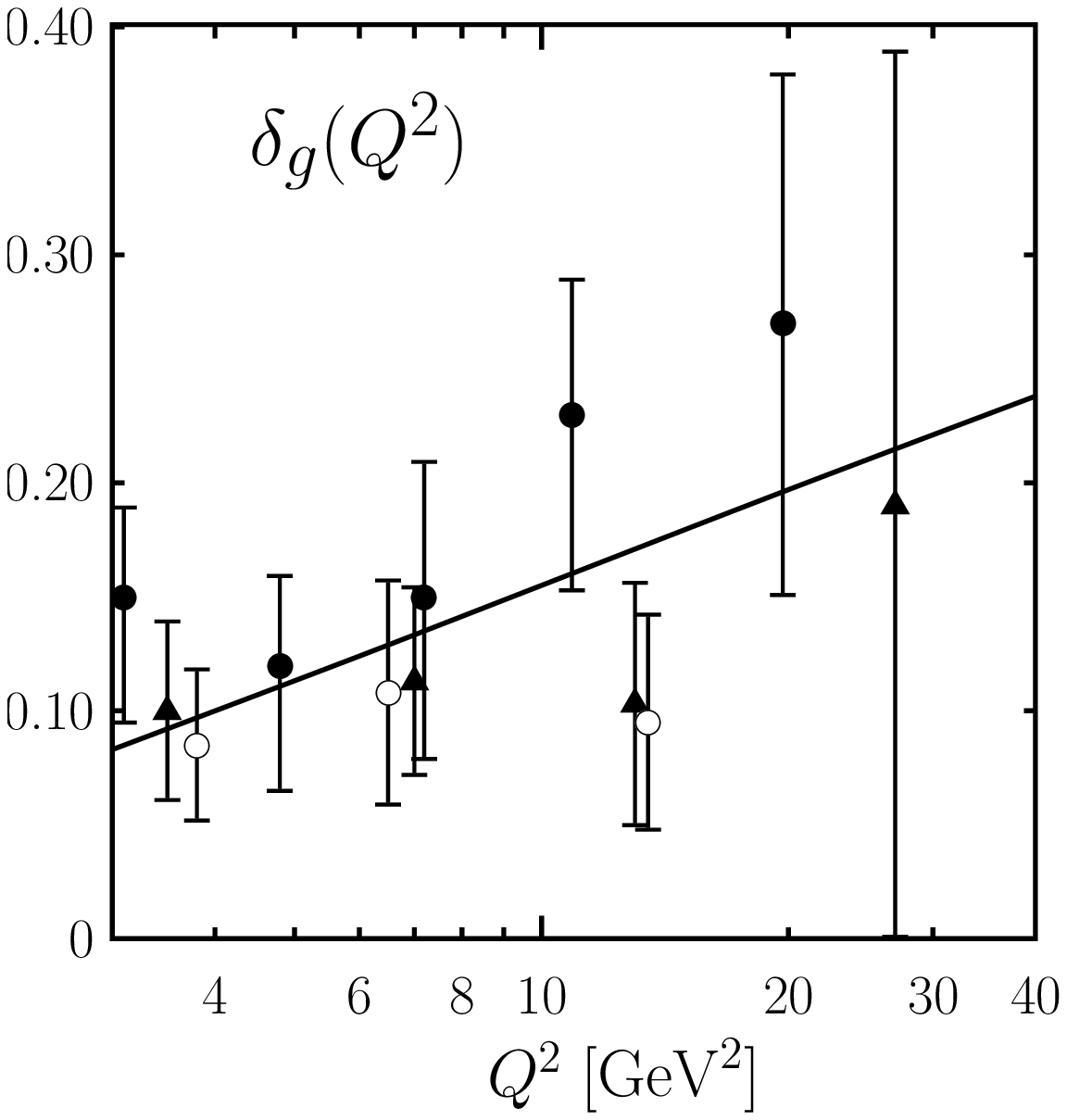} 
\end{center}
\caption{Left: Current gluon PDFs versus $\beta$ at a scale of 
$4\,\gev^2$ \ci{cteq6,MRST,alek,h1-pdf,zeus-pdf,GRV}. For the CTEQ6M 
solution only the band of Hessian errors is shown. Its width is 
typical of all PDFs. The central solid line represents the gluon PDF 
used in this work. Right: The intercept $\delta_g$ versus $Q^2$ at 
$W=75\,\gev$. The fit $\delta_g=0.10+0.06\ln{Q^2/Q^2_0}$ is compared 
to the HERA data for $\rho$ (solid circles \ci{h1}, triangles 
\ci{zeus98}) and $\phi$ electroproduction (open circles \ci{zeus05}).} 
\label{fig:gluon}
\end{figure}

In this section we are going to fix the parameters of our model and 
to present numerical results for the GPDs. For the evaluation of the
latter we have to choose a set of PDFs and to expand them according to
Eq.\ \req{pdf-exp}. Let us begin with the gluon PDF. The data used in 
current PDF analyses, e.g.\ Refs.\
\ci{cteq6,MRST,alek,h1-pdf,zeus-pdf,GRV}, do not constrain the parton 
distributions well for $\beta\lsim 10^{-2}$ at low $Q^2$
\ci{pumplin}. This is evident from Fig.\ \ref{fig:gluon} where
different versions of $\beta g(\beta)$ are displayed~\footnote{
   Here and in the following we denote the argument of the PDFs by 
   $\beta$ in parallel to the definition \req{function-h}
   in order to avoid confusion.}
at the scale $4\,\gev^2$; the deviations diminish with increasing scale. 
The uncertainties in the gluon PDF matter to vector meson 
electroproduction. Via Eqs.\ \req{DD}, \req{GPD-DD} and
\req{gluon-amp} they propagate to the cross section and lead to 
corresponding uncertainties there. In order to overcome this
deficiency we adjust the low-$\beta$ behavior of the gluon PDF in 
such a way that good agreement with the HERA data on $\rho$ and 
$\phi$ electroproduction is achieved. Using Eqs.\ \req{sigma}, 
\req{gluon-amp} and \req{xixi}, one readily obtains from the imaginary
part of the gluon contribution 
\be
\sigma_L \propto W^{\,4\,\delta_g(Q^2)}\,,
\label{sigmaL-W}
\ee
at fixed $Q^2$ and small $\xbj$; the real part does not affect the
energy dependence \ci{first}. This result allows for a determination 
of the intercept of the gluon trajectory. We stress that with 
$\delta_{\rm sea} =1+\delta_g$ the sea quark contribution leads to 
the same energy dependence of the cross section as the gluon and is
in so far included in \req{sigmaL-W}. For the large energy available 
at HERA the contributions from the valence quarks are negligible. 
 
\begin{figure}[t]
\begin{center}
\includegraphics[width=.30\textwidth,bb=109 385 426 720,%
clip=true]{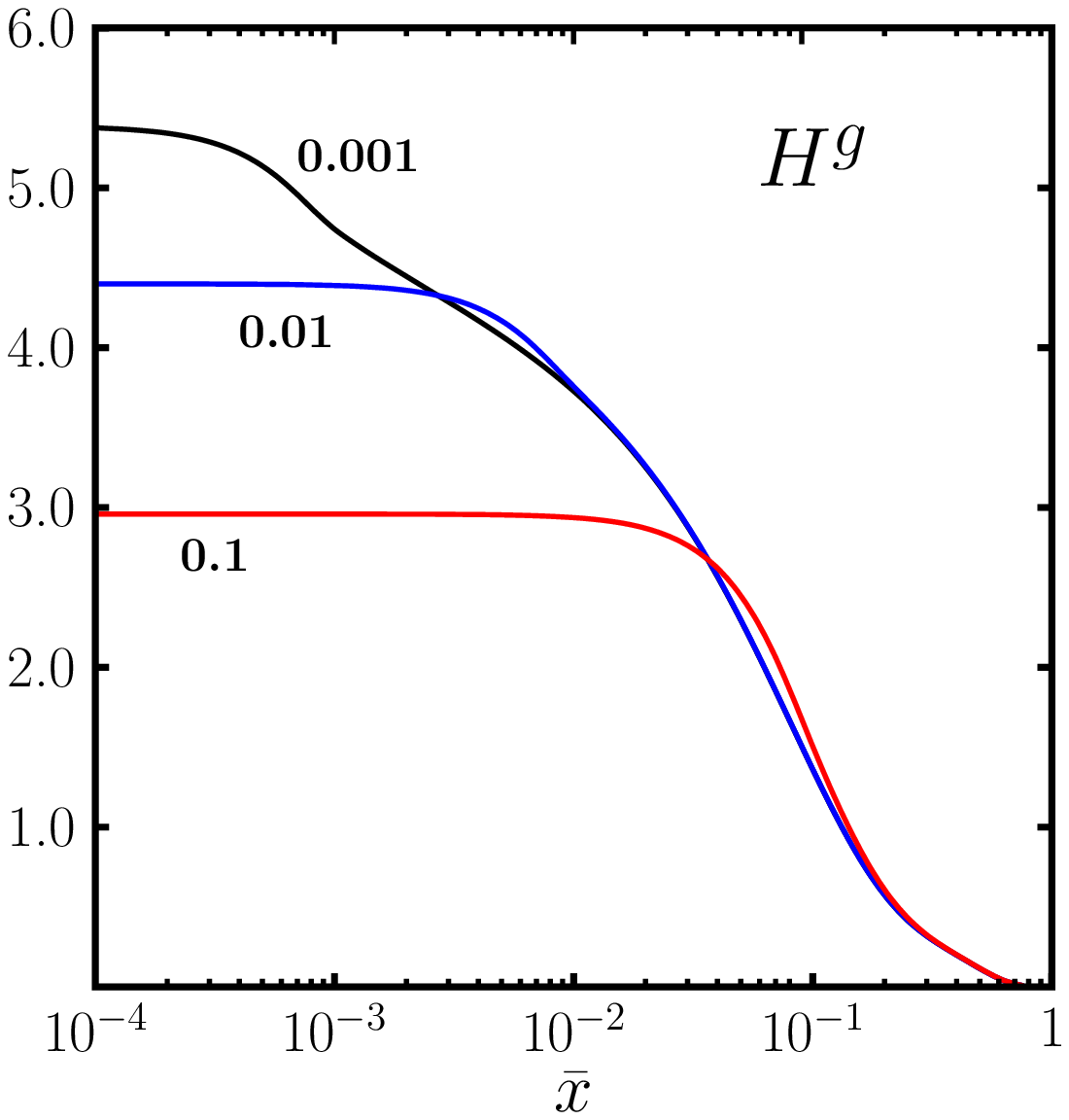} \hspace*{0.3cm}
\includegraphics[width=.30\textwidth,bb=145 337 461 673,%
clip=true]{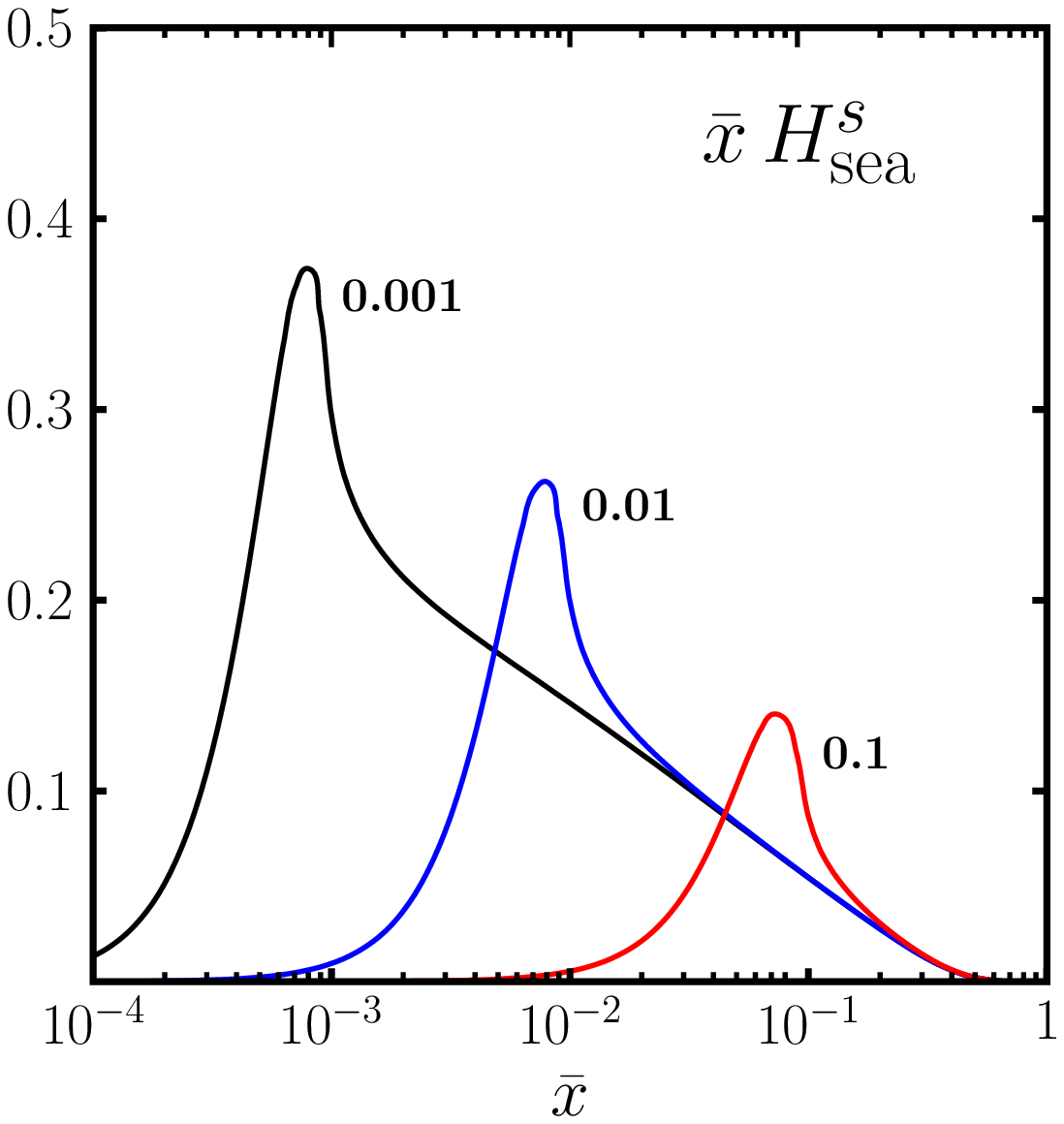} \hspace*{0.3cm} 
\includegraphics[width=.31\textwidth,bb=109 372 434 700,%
clip=true]{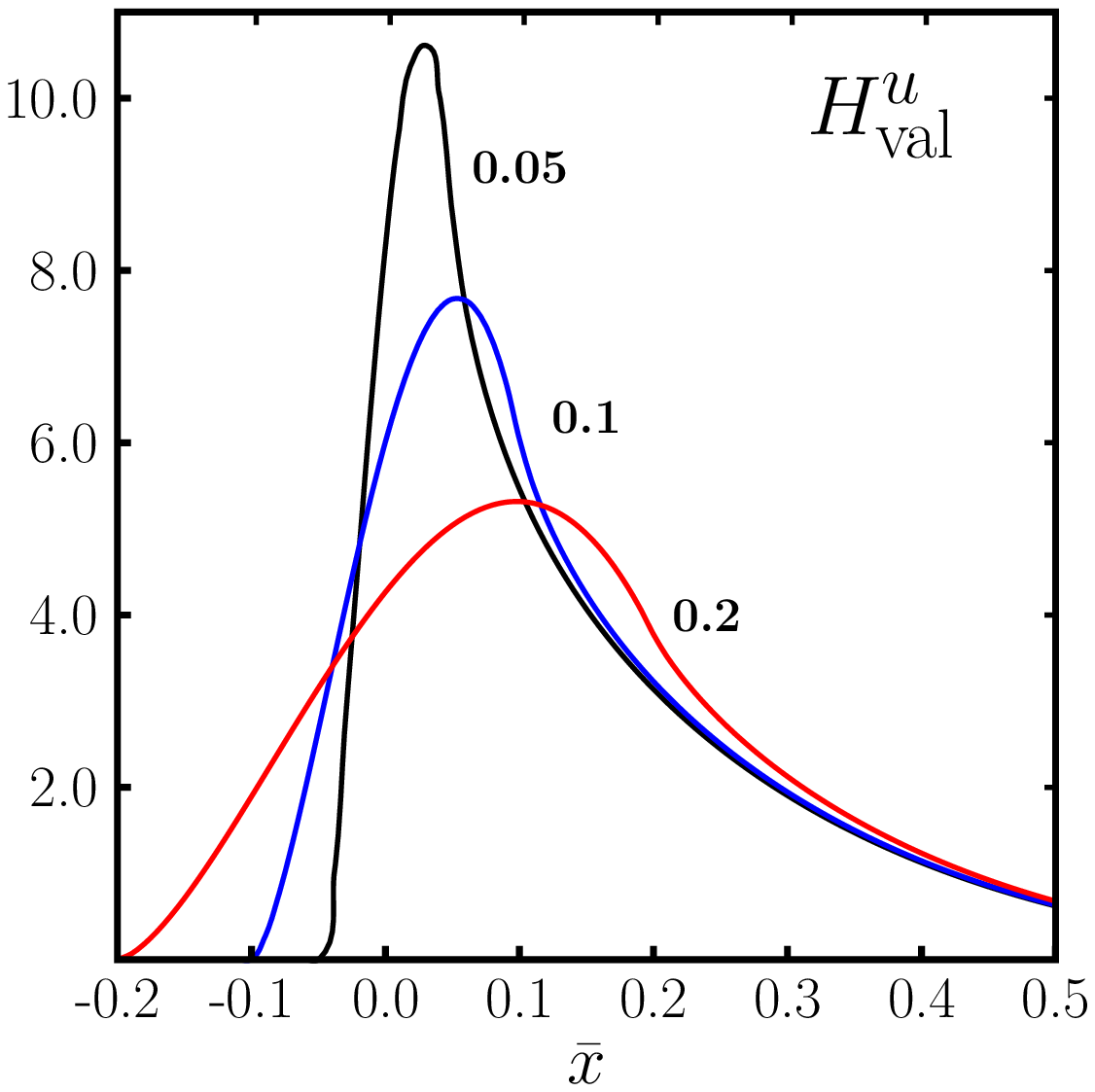}
\end{center}
\caption{The GPDs $H^g$ (left), $H^s_{\rm sea}$, multiplied by $\xb$, 
   (center), and $H_{\rm val}^u$ (right) at selected values of skewness.
   The GPDs are shown at $t'=0$ and at a scale of $4\,\gev^2$. } 
\label{fig:Hstrange}
\end{figure}

The power $\delta_g$ has been extracted from the HERA data 
\ci{h1,zeus98,zeus05} on the electroproduction cross section. The 
results are  shown in Fig.\ \ref{fig:gluon}. Since the ratio of the 
longitudinal and transverse cross section is only mildly energy
dependent the difference between the energy dependence of the full
cross section and of $\sigma_L$ is marginal. A straight-line fit to 
the HERA data is also shown in Fig.\ \ref{fig:gluon} and its
parameters are quoted in Tab.\ \ref{tab:par}. Keeping this result 
on $\delta_g$ subsequently fixed, the coefficients $c_{gj}$ in Eq.\ 
\req{pdf-exp} are fitted to the CTEQ6M gluon PDF \ci{cteq6} in the
range $4\,\gev^2 \leq Q^2 \leq 40\,\gev^2$ and $10^{-4} \leq \beta \leq
0.5$. The values obtained for the coefficients are given in Tab.\ 
\ref{tab:par} as well. The resulting fit is also shown 
in Fig.\ \ref{fig:gluon}. In the quoted range of $Q^2$ and
for $\beta$ the fit agrees very well with the CTEQ6M solution; it is 
always well inside the band of Hessian errors, see Fig.\
\ref{fig:gluon}. Larger values of $\beta$ are irrelevant to us since 
the contribution of the GPDs from the region $0.5 \lsim \beta$ to the 
real part of the amplitude is less than $0.5 \%$. From the gluon PDF 
just described we evaluate the gluon GPD with the help of Eqs.\ 
\req{GPD-exp} and \req{model2}. For a set of skewness values it is 
shown in Fig.\ \ref{fig:Hstrange}. In the context of the error
assessment to be executed below, we will comment on the implications
of the other PDF solutions.  
\begin{table*}
\renewcommand{\arraystretch}{1.4} 
\begin{center}
\begin{tabular}{|c|| c |c | c | c|}
\hline
         & gluon & strange & $u_{\rm val}$ & $d_{\rm val}$ \\
\hline
$\delta$ & $0.10+0.06\,L$ & $1+\delta_g$ & 0.48 & 0.48 \\ 
$c_0$    & $2.23+0.362\,L$ & $\phantom{-}0.123+0.0003\,L$
         & $1.52+0.248\,L$& $ 0.76+0.248\,L$ \\
$c_1$    & $\phantom{-}5.43-7.00\,L$ & $-0.327-0.004\,L$ & 
                                $2.88-0.940\,L$ & $\phantom{-}3.11-1.36\,L$ \\
$c_2$    & $-34.0+22.5\,L$ & $\phantom{-}0.692-0.068\,L$& $-0.095\,L$
         & $-3.99+1.15\,L$ \\
$c_3$    & $\phantom{-}40.6 -21.6\,L$ & $-0.486+0.038\,L$ & $0$ & $0$ \\
\hline
\end{tabular}
\end{center}
\caption{The parameters appearing in the expansion \req{pdf-exp} of 
the PDFs ($L=\ln{Q^2/Q^2_0}$, $Q^2_0=4\,\gev^2$). The expansion
provide fits to the CTEQ6M PDFs \ci{cteq6} in the range 
$10^{-2} \leq \beta \leq 0.5$ and $ 4\,\gev^2 \leq Q^2 \leq
40\,\gev^2$. The powers $\delta$ are kept fixed in the fits.}
\label{tab:par}
\renewcommand{\arraystretch}{1.0}   
\end{table*} 

A power $\delta_g$ rising with $Q^2$, is untypical for the
Regge-pole model. Even more important, $\delta_g>0$ leads to a cross
section that increases as a power of the energy, see Eq.\
\req{sigmaL-W}. At very high energies, it will therefore violate the
Froissart bound and, hence, unitarity. Obviously, there must occur a
saturation at some scale of $Q^2$ that will limit the rise of the 
gluon PDF and GPD and will restore unitarity. In other words, the 
description of diffraction by a Pomeron-type pole with an intercept, 
$\alpha_g(0)$, larger than unity is to be considered as an effective 
parameterization that holds in a finite although possibly large
range of energy \ci{sandy}. Unitarity will ultimately force the
generation of a series of shielding cuts \ci{oehme} that will prevent
the violation of the Froissart bound. An effective parameterization 
may have parameters that dependent on the process and on the kinematics. 
 
For the slope of the gluon trajectory  we take the value 
$\alpha_g'=0.15\,\gev^{-2}$ which is slightly smaller than that of the
usual soft Pomeron \ci{sandy} but agrees with the value observed in 
photoproduction of the $J/\Psi$ \ci{zeus02} and other vector mesons
\ci{ivanov04}. Small values of the gluon and soft Pomeron slopes are 
required since the diffraction peaks show little shrinkage. 

Using $\delta_{\rm sea}=1+\delta_g$ we perform an analogous fit of the
strange quark CTEQ6M distribution (assuming $s(\beta)=\bar{s}(\beta)$). 
The parameters are quoted in Tab.\ \ref{tab:par} too and the resulting 
GPD $H^s_{\rm sea}$ constructed through the double distribution
\req{DD} is shown in Fig.\ \ref{fig:Hstrange}. For the CTEQ6M PDFs 
\ci{cteq6} the $\bar{u}$ and $\bar{d}$ distributions at low $\beta$
are very close to each other and enhanced by an approximately 
$Q^2$-dependent but $\beta$-independent factor $\kappa_s(Q^2)$ as 
compared to the strange quark PDF. In an attempt to keep the GPD model 
simple we therefore assume
\be
H^u_{\rm sea} \= H^d_{\rm sea} \= \kappa_s\,H^s_{\rm sea}\,,
\label{eq:sea}
\ee
where the flavor symmetry breaking factor is parameterized as
\be
\kappa_s \= 1+ 0.68/(1+0.52\,\ln{Q^2/Q^2_0})\,,
\label{eq:kappas}
\ee
as obtained from a fit to the CTEQ6M PDFs. 
Eq.\ \req{eq:sea} is a simplification which as one may object, is
unjustified given the high level of accuracy the current PDF solution
have reached. However we are constructing model GPDs which implies a
theoretical uncertainty of unknown strength. It seems premature in the 
present state of the art to transfer the full complexity of the
current PDFs to the model GPDs. It is not probed by the present data 
on vector meson electroproduction~\footnote{
Since as yet there are only data on $\rho$ production available but
not on $\omega$ only the combination $e_u\,H^u_{\rm sea} - e_d\,H^d_{\rm sea}$
is probed.}
and would rather confuse than elucidate the physical interpretation. 

For the valence trajectory we adopt standard soft physics parameters, 
$\alpha_{\rm val}(0)=0.48$ and $\alpha_{\rm val}'=0.9\,\gev^{-2}$. 
This is in fair agreement with the low-$x$ behavior of the valence 
quark PDFs at low factorization scale (up to at least $40\,\gev^2$) 
within errors. Only a very mild, negligible effect of evolution is to 
be observed for these parameters. Keeping again the intercept fixed we 
fit the expansion \req{pdf-exp} to the $u$ and $d$ valence quark 
distributions of the CTEQ6M solution and evaluate the correponding
GPDs from Eqs.\ \req{GPD-exp} and \req{model1}. The obtained 
parameters are quoted in Tab.\ \ref{tab:par} and $H^u_{\rm val}$ at
$t'=0$ is displayed in Fig.\ \ref{fig:Hstrange}. In contrast to the
sea quark GPDs where the $n=1$ model is to be rejected because of its 
very strong skewing effect (and because of the Pomeron-type
interpretation), the skewing effect for the valence quarks is much weaker. 

\begin{figure}[t]
\begin{center}
\includegraphics[width=.38\textwidth,bb= 125 363 483 714, clip=true]
{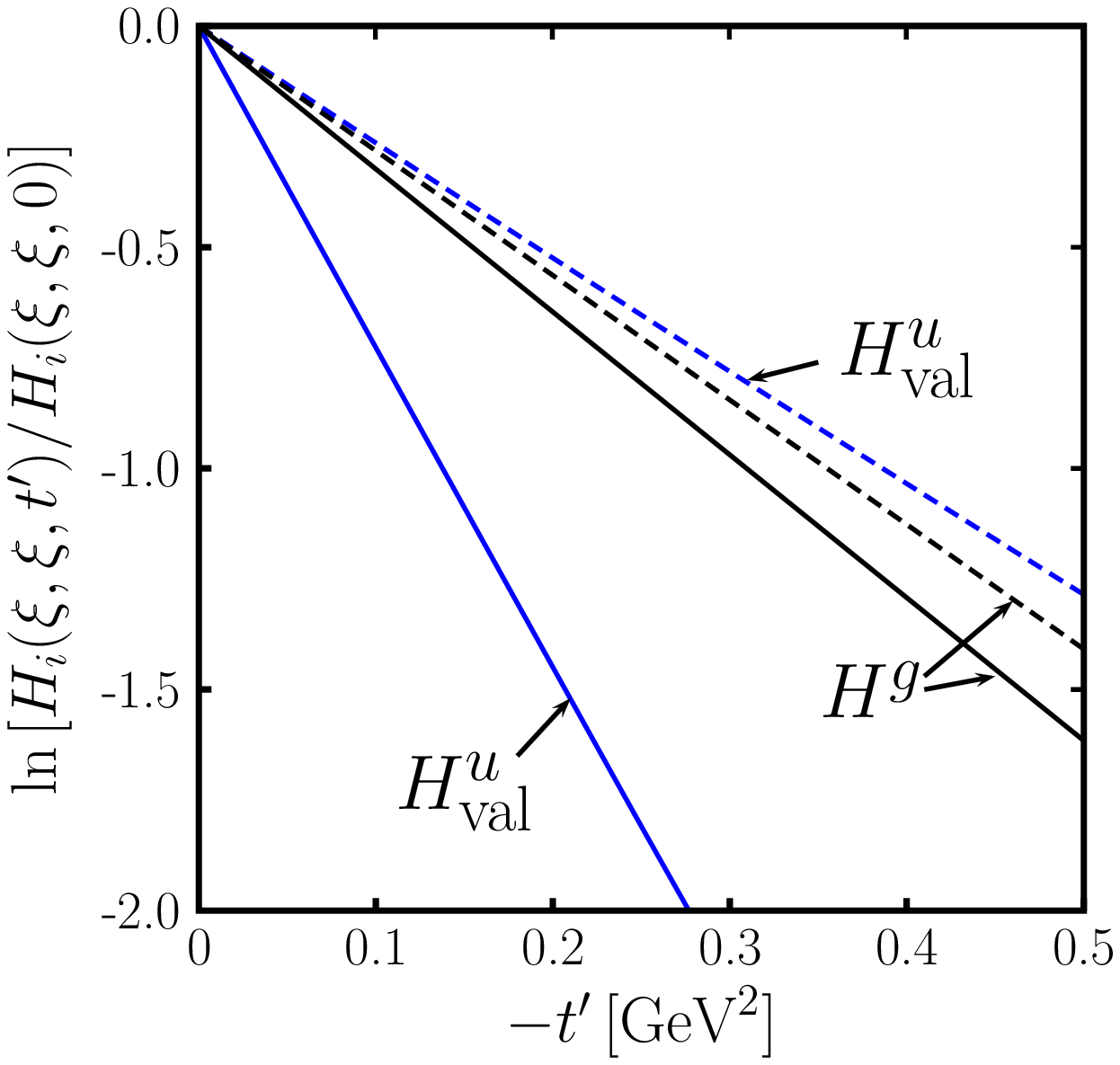}\hspace*{1.0em}
\includegraphics[width=.46\textwidth,bb= 28 355 543 767,clip=true]
{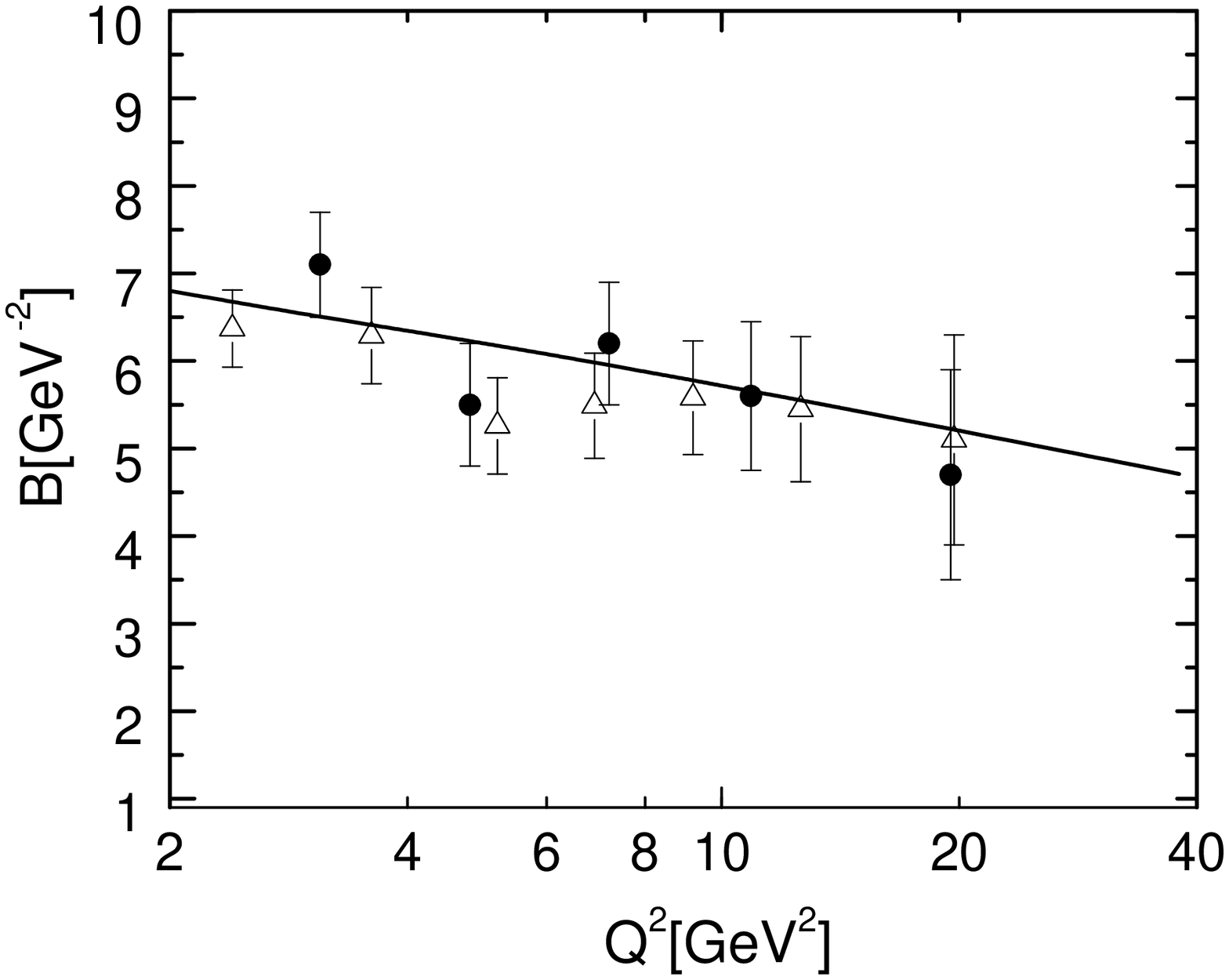}
\end{center}
\caption{Left: The $t'$ dependence of the gluon and the 
  $u$-valence GPDs at $W=75\,\gev,\; Q^2=4\,\gev^2$ 
  (solid lines) and $W=5\,\gev\,\; Q^2=3\,\gev^2$ (dashed lines). 
   Right: The slope $B_V$ of the differential cross section for $\rho$ 
  and $\phi$ electroproduction versus $Q^2$ at $W=75\,\gev$. Data are 
  taken from Ref.\ \ci{h1} (solid circles) and \ci{zeus05} (open 
  triangles). The solid line represents the fit \req{slope},
  \req{constant-slope} for the case of the $\phi$.}
\label{fig:slope}
\end{figure}

Now having specified the parameterization of the Regge trajectories we
turn to their determination of the residues, i.e.\ of the slope
parameters $b_i$. As we  repeatedly mentioned vector meson 
electroproduction behaves diffractively, i.e.\ its differential cross 
section decrease exponentially with $t'$. In the handbag approach this
behavior is to be incorporated in the gluon GPD. Although its $t'$ 
dependence appears to be more complicated than an exponential as is
for instance seen from Eq.\ \req{model2} or \req{xixi}, this is not
the case in reality. They actually behave as exponentials as can be
seen from Fig.\ \ref{fig:slope} where we display $H^g(\xi,\xi,t')$ 
versus $t'$ for selected values of $W$ and $Q^2$. As a consequence of 
the smallness of $\alpha'_g$ the effective slope of the gluon GPD 
falls together with the Regge exponential to a very high degree of 
accuracy, for the sea quark GPD the situation is similar. Hence,
because of the repeatedly mentioned dominance of the imaginary part of 
the gluon contribution at HERA kinematics the slope of the
differential cross section is given by
\be
B_V \simeq 2 b_g + 2 \alpha'_g \ln{\frac{1+\xi}{2\xi}}\,,
\ee
in our model GPDs. Insertion of Eqs.\ \req{xbj} and \req{xi-xbj}
translates this relation into 
\be
B_V \= 2 b_g + 2 \alpha'_g \ln{\frac{W^2+Q^2}{Q^2+m_V^2}}\,.
\label{slope}
\ee
High-energy data for the $t'$ dependence only exist for the
unseparated cross section $d\sigma=d\sigma_T +
\varepsilon\,d\sigma_L$ where $\varepsilon$ is the ratio of
longitudinal to transversal polarization of the virtual photon. 
Ignoring possible differences between the slopes of the transverse 
and longitudinal cross sections (which can only emerge from 
the subprocess amplitudes) we fit $b_g$ against the HERA data for 
$\rho$ \ci{h1} and $\phi$ production \ci{zeus05}. We find that the 
experimental slope parameters are well described by
\be
b_g \= b_{\rm sea} \= 2.58\, \gev^{-2} + 0.25\,\gev^{-2}\,
  \ln{\frac{m^2}{Q^2+m^2}}\,,
\label{constant-slope}
\ee
see Fig.\ \ref{fig:slope}. It is to be stressed that the term $\propto 
\alpha_g'$ in Eq.\ \req{slope} is a consequence of the Regge behavior
while the log term in Eq.\ \req{constant-slope} is an ansatz.
The $\rho$ and $\phi$ slopes practically fall together at HERA 
energies; there are only minor differences at low $Q^2$ which are not 
shown in Fig.\ \ref{fig:slope}. 

The zero-skewness limit of the valence quark GPDs reads
\be
H^q_{\rm val}(\xb,\xi=0,t')\={\rm e}^{(b_{\rm val}+\alpha'_{\rm
    val}\,\ln(1/\xb))\,t'}\,q_{\rm val}(\xb)\,,
\ee
as readily follows from \req{DD} and \req{GPD-DD}. This is very close
to an ansatz advocated for in Ref.\ \ci{DFJK4} in order to extract the 
zero-skewness GPDs from the nucleon form factor data (see also Ref.\ 
\ci{rad-van}). Our Regge 
exponential appears as the small $\xb$ approximation of the
exponential exploited in Ref.\ \ci{DFJK4}. In the range 
$0\lsim \xb \lsim 0.15$ the difference between both the exponentials
is less than $10\%$. The comparison with the zero-skewness analysis 
further reveals  that the slope of the Regge trajectory suffices to 
specify the $t'$ dependence of the valence quark GPDs. 
Consequently we set $b_{\rm val}$ equal to zero. It is 
also checked by us that the valence quark GPDs \req{GPD-exp}, 
\req{model1} respect the sum rule for the Dirac form factor at any 
value of skewness; small deviations between the form factor data and 
the sum rules evaluated from our GPDs occur at larger $t'$ and amount
to about $3\%$ at $t=-0.6\,\gev^2$.

Effectively the valence quarks GPDs behave as exponentials in $t'$ too,
see Fig.\ \ref{fig:slope}. The decisive difference is however
that the valence quark GPDs show strong shrinkage due to the large
value of $\alpha'_{\rm val}$. While at HERMES kinematics the effective
slopes of the gluon and the valence quark GPDs are similar, is the
latter much larger at HERA kinematics.

Last not least we have to specify the meson wavefunction occurring in
Eq.\ \req{mod-amp}. As in our previous work \ci{first} and in other
applications of the modified perturbative approach, e.g.\
\ci{jak93,DaJaKro:95}, we use a Gaussian \wf{}
\be
\Psi_V(\tau,\vk) \= 8\pi^2\sqrt{2N_c} f_V a_V^2 
                        {\rm exp}{[-a_V^2\vk^2/(\tau\bar{\tau})]}\,.
\label{wavefunction}
\ee
Transverse momentum integration of it leads to the associated
distribution amplitude which represents the soft hadronic matrix
element entering calculations in the collinear factorization
approach. Actually, the \wf{} \req{wavefunction} leads to the
so-called asymptotic form of the meson distribution amplitude
\be
\Phi_{\rm AS} \= 6 \tau \bar{\tau}\,.
\label{meson-da}
\ee
Its $1/\tau$ moment occurs in the leading-twist result \req{lt} and
acquires the value 3. The transverse size parameter $a_V$ is
considered as a free parameter fitted to the data on the
integrated cross sections $\sigma_L$ for $\rho$ and $\phi$
production, see next section. It can be varied within a certain range
of values determined by the requirement that the corresponding 
r.m.s. $\vk$ being related to the transverse size
parameter by
\be
\langle k_\perp^2 \rangle^{1/2} \= \big[\sqrt{10}\,a_V\big]^{-1}\,,
\ee
acquires a plausible value consistent with our assumption of taking into
account only the transverse momenta of the quarks forming the meson. 
A possible evolution of the transverse size parameter is ignored.

\section{Results for the longitudinal cross sections}
\label{sec:results}
The full amplitudes for vector meson electroproduction \req{gluon-amp}, 
\req{quark-amp} are a coherent superposition of contributions from the 
various quark flavors and from the gluon. In order to shed light on
the relative importance of the various terms we quote the
leading-twist result
\ba
{\cal M}_\phi &=& e \frac{8\pi\als}{N_cQ}\, f_\phi \langle
                         1/\tau\rangle_\phi\,
     \frac{-1}{3}\,\left\{\frac1{2\xi}I_g + C_F\,I_{\rm sea}\right\}\,, \nn\\
{\cal M}_\rho &=& e \frac{8\pi\als}{N_cQ}\, f_\rho \langle
                         1/\tau\rangle_\rho\,
   \frac1{\sqrt{2}}\,\left\{\frac1{2\xi}I_g + \kappa_s\, C_F\, I_{\rm sea}
      +\frac13\, C_F\,I^u_{\rm val} + \frac16\, C_F\,I^d_{\rm val}\right\}\,,
\nn\\
{\cal M}_\omega &=&  e \frac{8\pi\als}{N_cQ}\, f_\omega \langle
                         1/\tau\rangle_\omega\,
           \frac1{3\sqrt{2}}\,\left\{\frac1{2\xi}I_g +
           \kappa_s\, C_F\, I_{\rm sea}
          + C_F\,I^u_{\rm val} - \frac12\,C_F\,I^d_{\rm val}\right\}\,.
\label{eq:lt-amp}
\ea
The integrals in  Eq.\ \req{eq:lt-amp} read
\ba
I_g &=& 2\int_0^1 d\xb \frac{\xi
      H^g(\xb,\xi,t')}{(\xb+\xi)(\xb-\xi+i\eps)}\,,\nn\\
I_{\rm sea} &=& 2\int_0^1 d\xb \frac{\xb H^s_{\rm sea}(\xb,\xi,t')}
           {(\xb+\xi)(\xb-\xi+i\eps)}\,,\nn\\
I^a_{\rm val} &=& 2\int_{-\xi}^1 d\xb \frac{\xb H^a_{\rm val}(\xb,\xi,t')}
           {(\xb+\xi)(\xb-\xi+i\eps)}\,.
\label{integrals}
\ea
The imaginary parts of these integrals are just $-\pi H_i(\xi,\xi,t')$
and for $\xi\to 0$ they exhibit typical Regge phases as a consequence
of the analytic structure of the integrals and the symmetries of the
GPDs \req{symmetry}.

Within the modified perturbative approach the amplitudes have the same
structure as in Eq.\ \req{eq:lt-amp}. Only the integrals
\req{integrals} are much more complex, they do not factorize into 
products of integrals over the wave functions and such over the product
of GPDs and propagators. Despite this the suppressions induced by the 
modified perturbative approach do not change much the relative
strengths of the various contributions. One may therefore get an quick 
insight into the relative strength of the various terms from Eq.\ 
\req{eq:lt-amp}. 

\begin{figure}[t]
\begin{center}
\includegraphics[width=.45\textwidth, bb=27 341 545 704,clip=true]
{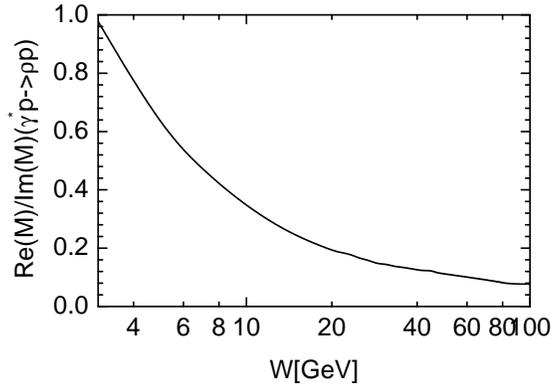}
\end{center}
\caption{Real over imaginary part of the full amplitude for $\rho$
  production versus $W$ at $Q^2=4\,\gev^2$ and $t'=0$.}
\label{fig:phase}
\end{figure}
Numerical evaluation of the amplitudes \req{gluon-amp} and
\req{quark-amp} reveals that, for skewness less than about 0.01, the 
gluon and sea contributions are dominantly imaginary while, for 
$\xi \simeq 0.1$, their real parts are nearly as large as their
imaginary part. The valence quark contribution behaves oppositely. 
The energy dependence of ratio of the real and imaginary parts of the 
full $\rho$ production amplitude is shown in Fig.\ \ref{fig:phase} at 
$Q^2=4\,\gev^2$. A remark concerning the values of $\als$ is in
order. As an inspection of our handbag amplitude reveals almost the
entire contribution is accumulated in a comparatively narrow region of
$\als$. For instance, at $Q^2=4\,\gev^2$ and $W=5\,\gev$ $90\%$ of the
amplitude comes from the range $0.3\lsim \als \lsim 0.5$ with a mean
value of about 0.4. Hence, our handbag approach is theoretically
self-consistent in so far as contributions from soft regions where
$\als$ is larger than, say, 0.6 and where perturbation theory breaks
down, are strongly suppressed. 
    
\begin{figure}[pht]
\begin{center}
\includegraphics[width=.41\textwidth, bb=33 298 552 690,clip=true]
{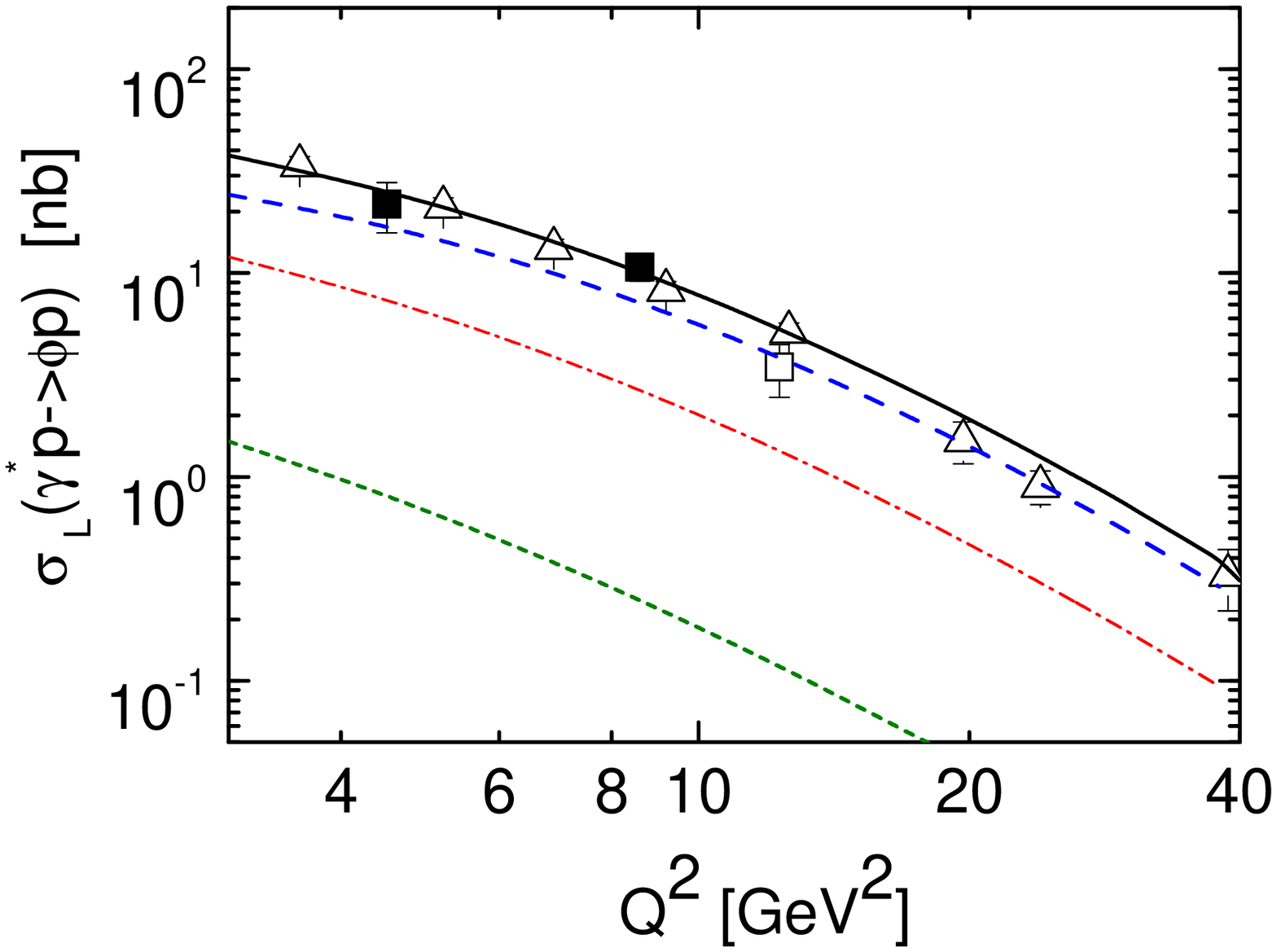}\hspace*{1.0em}
\includegraphics[width=.39\textwidth,bb= 33 303 533 700,clip=true]
{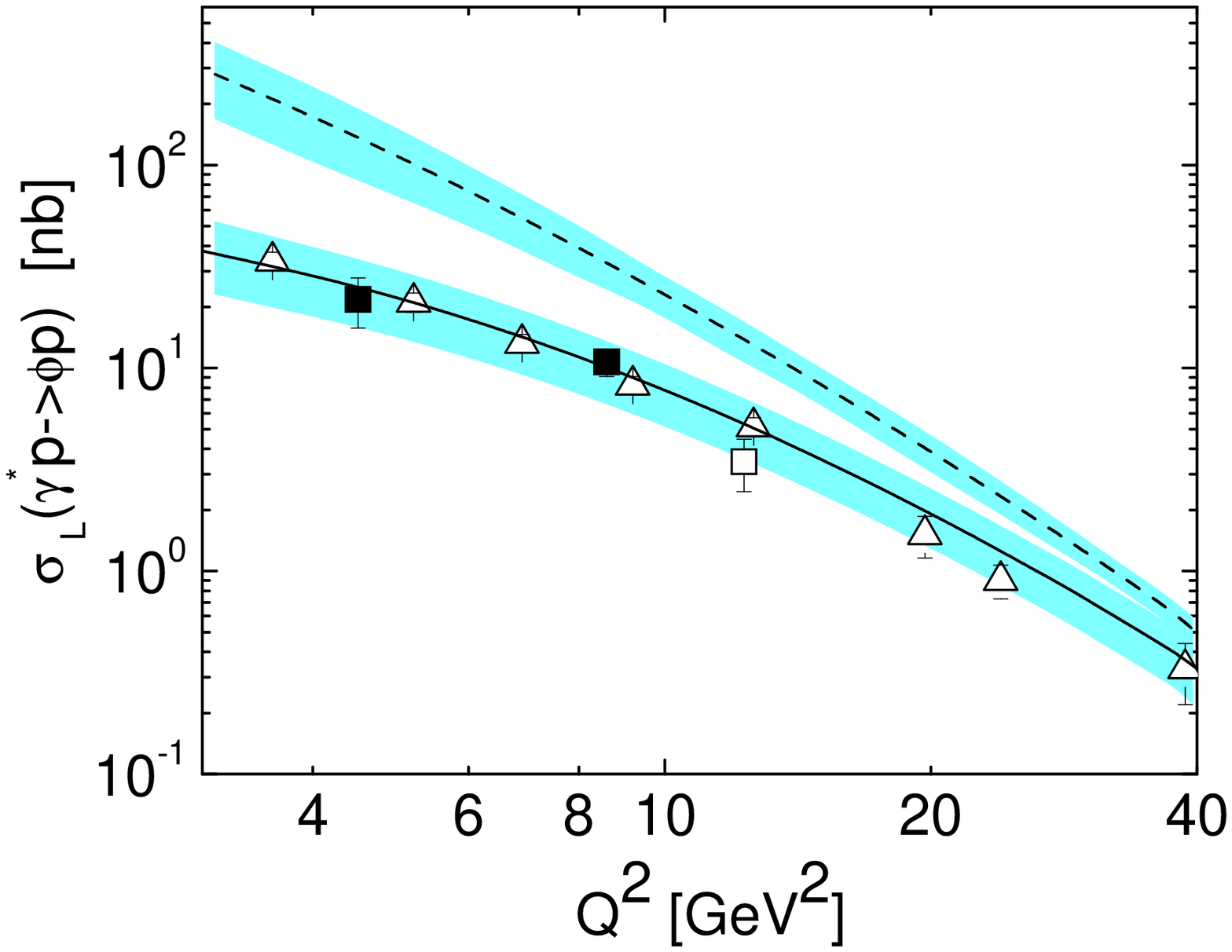}
\end{center}
\caption{The longitudinal cross section for $\phi$ production at
  $W=75\,\gev$. Data are taken from \ci{zeus05} (open triangles),
  \ci{adloff} (solid squares) and \ci{zeus96} (open squares). Left: 
  Full (dashed, dash-dotted, dotted) line represents the handbag predictions 
  for the cross section (gluon, gluon-sea interference, sea contribution).
  Right: Predictions for the cross section with error bands 
  resulting from the Hessian errors of the CTEQ parton distributions 
  (full line) and compared to the leading-twist result (dashed line).}
\label{fig:sigmaL-Q}
\begin{center}
\includegraphics[width=.40\textwidth, bb=31 310 533 700,clip=true]
{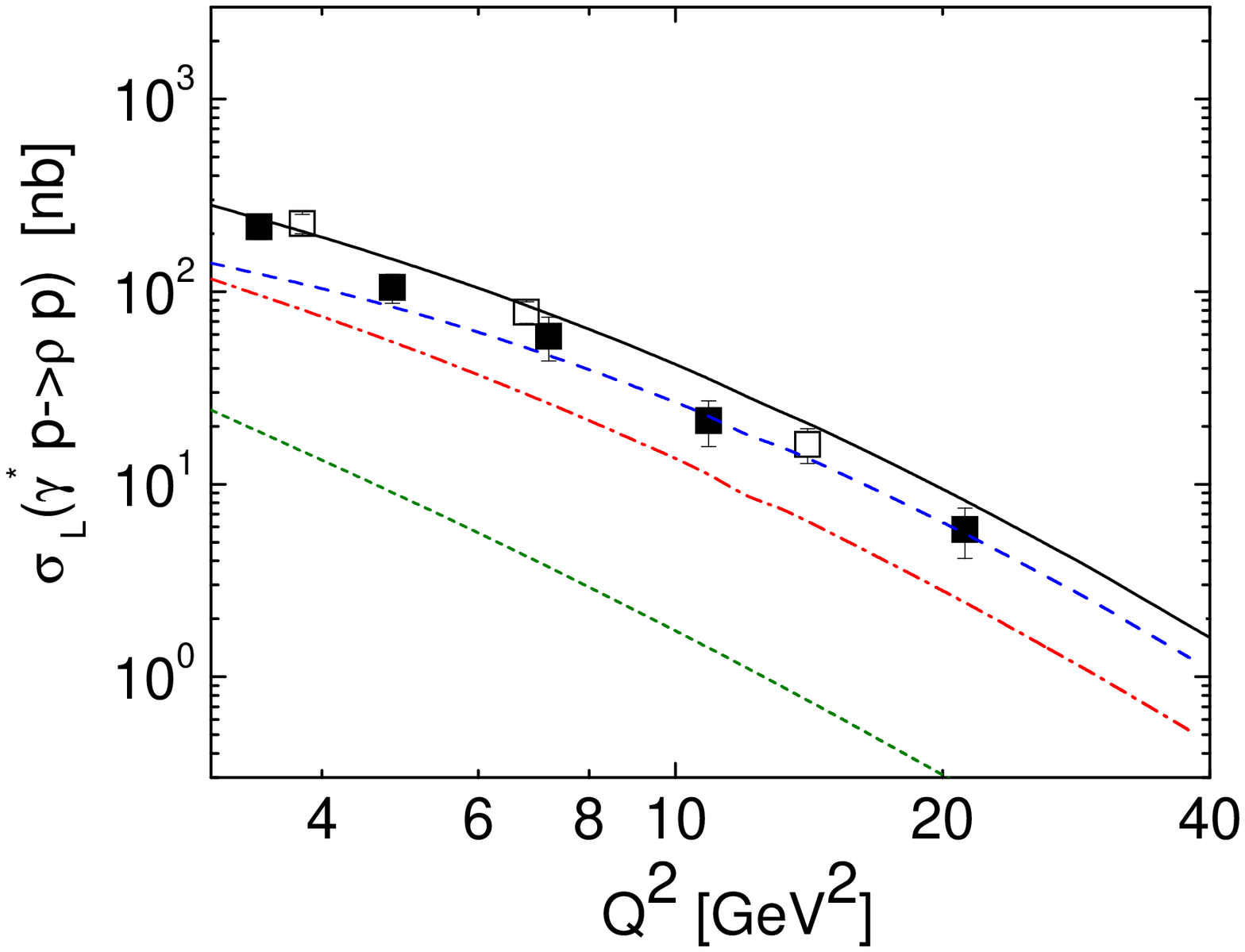}
\hspace*{1.0em}
\includegraphics[width=.40\textwidth,bb= 37 313 534 700,clip=true]
{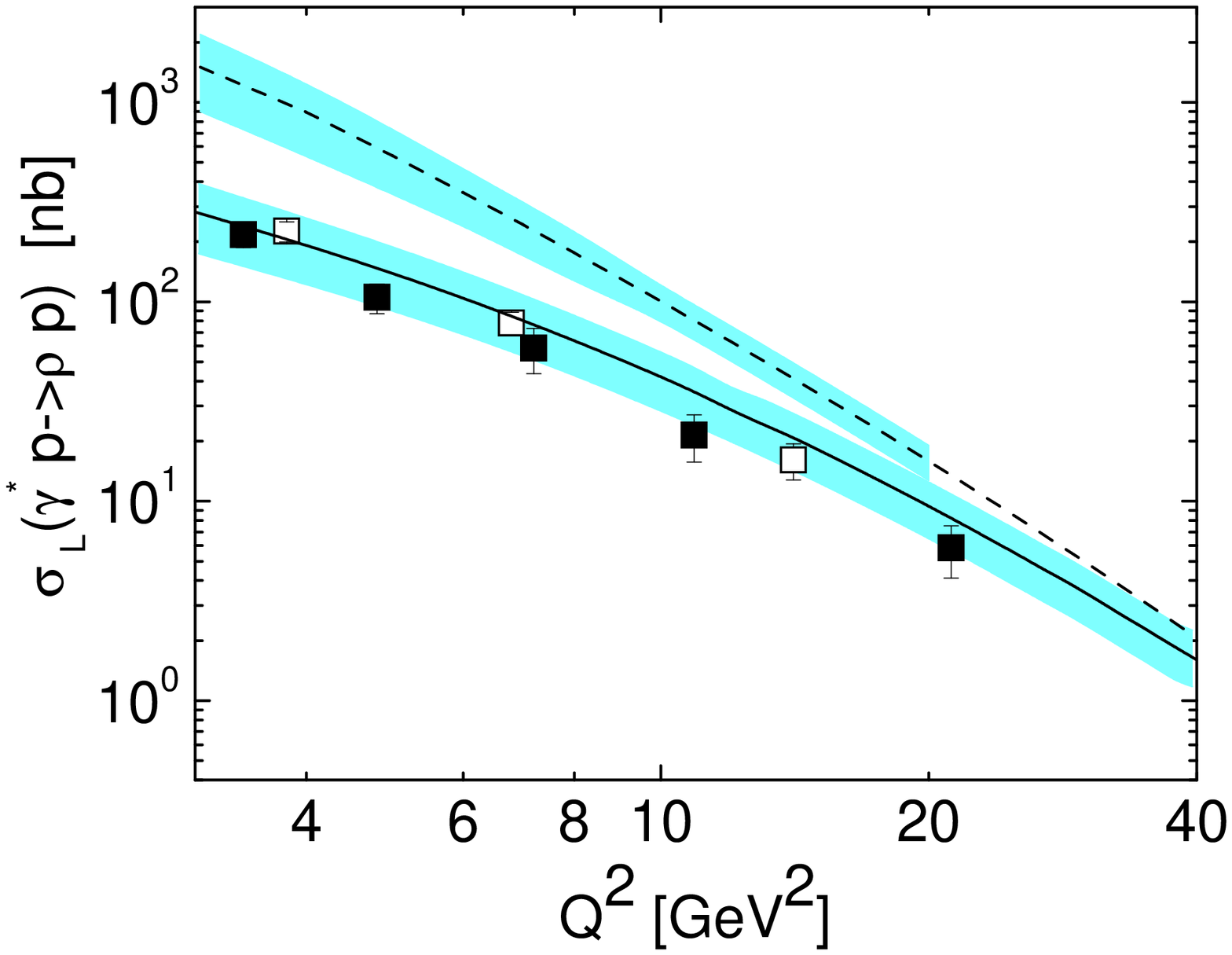}
\end{center}
\caption{The longitudinal cross section for $\rho$ production at
  $W=75\,\gev$. Data are taken from \ci{h1,h196} (solid squares) and
  \ci{zeus98} (open squares). For other notations cf.\ Fig.\
  \ref{fig:sigmaL-Q}.}
\label{fig:sigmaL-Q-rho}
\end{figure} 
We are now ready to present our results for vector meson
electroproduction. They are obtained by adjusting the transverse 
size parameters $a_V$ appropriately. The best fits provide the values 
\be
a_\phi \= 0.70\,\gev^{-1}\,, \qquad \qquad a_\rho\=0.75\,\gev^{-1} \,.
\label{eq:trans-size}
\ee
The corresponding r.m.s. $\vk$ is about $500\, \mev$. This value is 
more than a factor of 2 larger than the corresponding value for 
the quarks inside the proton and is in so far consistent with our 
assumption of taking into account only the transverse momenta of 
the quarks forming the meson. In Figs. \ref{fig:sigmaL-Q} and 
\ref{fig:sigmaL-Q-rho} we compare our results for $\phi$ and $\rho$ 
production to the HERA data~\footnote{
If not quoted explicitely in the experimental papers the longitudinal
cross section is evaluated by us from information given therein. 
Statistical and systematical errors are added in quadrature. If 
necessary data are rescaled in $W$ using Eq.\ \req{sigmaL-W} or $Q^2$ 
exploiting the handbag predictions for the dependence on it.} 
\ci{h1,zeus98,zeus05,adloff,zeus96,h196}. In the left panels of these
figures we show the decomposition of the cross sections into the
various contributions from the gluon and the quarks. The gluon
contribution is dominant but the corrections from the gluon-sea 
interference amount to about $25\, (50)\%$ for $\phi$ ($\rho$)
production; the valence quark contribution (including its interference 
with the gluon and the sea) to the $\rho$ production cross section  
is tiny and can be neglected. The larger sea quark contribution
for $\rho$ production follows from the flavor symmetry breaking
factor $\kappa_s$. In the right panels of Figs. \ref{fig:sigmaL-Q} and 
\ref{fig:sigmaL-Q-rho}  we display in addition to our full results
their uncertainties due to the Hessian errors of the CTEQ6 PDFs and 
for comparison the leading-twist result evaluated from the asymptotic
distribution amplitude \req{meson-da}. Results for the cross
sections evaluated from sets of PDFs other than CTEQ6 also fall into
the error bands in most cases (an exception is set for instance by the
PDFs determined in Ref.\ \ci{GRV}) provided these PDFs are treated
in analogy to the CTEQ6M set, i.e.\ they are fitted to the 
expansion \req{pdf-exp} by forcing them to behave Regge-like with 
powers $\delta_i$ as described above, and if necessary readjusting 
the transverse size parameters. Straight-forward evaluation of the 
GPDs from the various sets of PDFs and fixed transverse-size
parameters lead to cross sections which differ markedly stronger
than the error bands indicate. For examples see Ref.\ \ci{kugler}.
The results obtained with the 
modified perturbative approach are in remarkable agreement with the 
HERA data while the leading-twist results are clearly in excess to 
experiment with a tendency however of approaching the data and the 
predictions from the modified perturbative approach at 
$Q^2 \simeq 40\,\gev^2$. This in turn tells us that the effect of the 
transverse quark degrees of freedom in combination with the Sudakov 
suppressions become small for such values of $Q^2$ while being very 
important at lower $Q^2$. Pertinent observations have also been made 
by Ivanov {\it et al.} \ci{iva04}. In their next-to-leading order 
leading-twist calculation of vector meson electroproduction large 
perturbative logs occur which partly cancel the leading-order term 
bringing the leading-twist result closer to experiment. These logs 
are included in the Sudakov factor \req{eq:sudakov}. 

\begin{figure}[t]
\begin{center}
\includegraphics[width=.45\textwidth, bb=32 315 529 710,clip=true]
{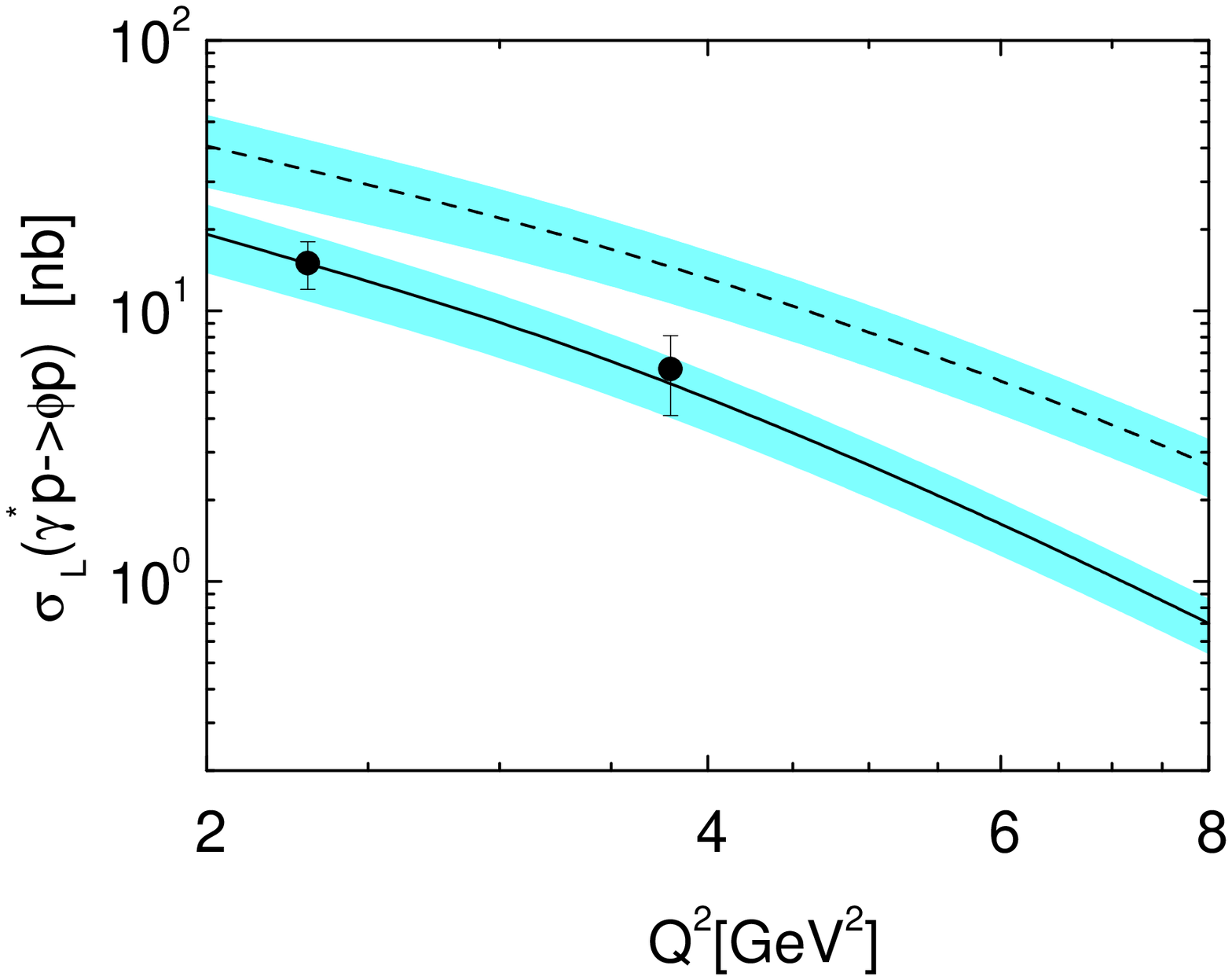}\hspace*{1.0em}
\includegraphics[width=.44\textwidth,bb= 40 316 528 700,clip=true]
{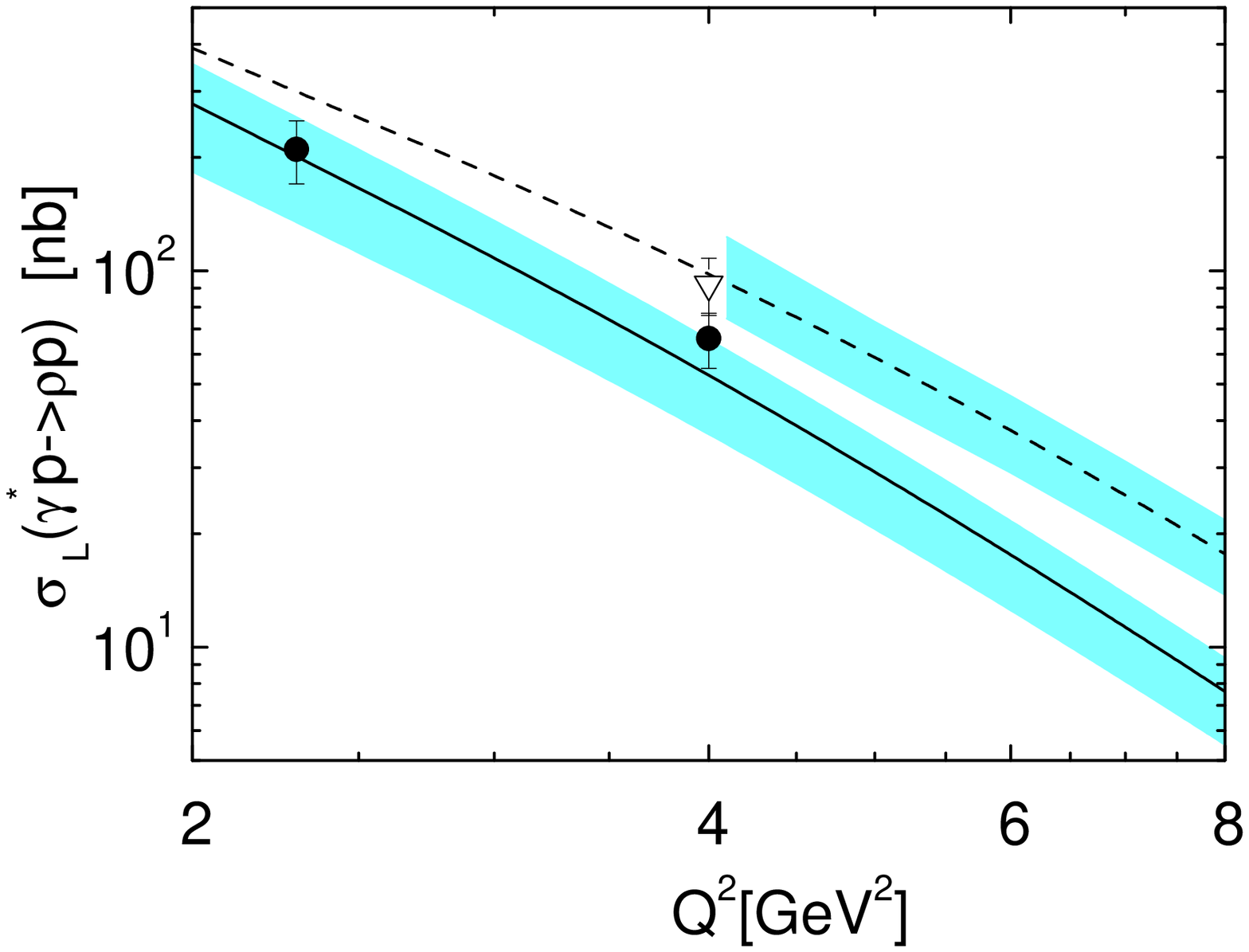}
\end{center}
\caption{The longitudinal cross section for $\phi$ (left) and $\rho$
  (right) production versus $Q^2$ at $W=5$ (solid line) and $10\,\gev$ 
  (dashed line). Data at $W=5\,\gev$, shown as solid circles, are
  taken from HERMES \ci{HERMES-prel} (for $\phi$, preliminary)  and \ci{hermes-rho} 
  (for $\rho$). The open triangle presents the E665 \ci{e665} data 
  point at $W=10\,\gev$. For other notations cf.\  Fig.\ \ref{fig:sigmaL-Q}.}
\label{fig:sigmaL-lowW}
\end{figure}
\begin{figure}[t]
\begin{center}
\includegraphics[width=.45\textwidth, bb=32 317 542 685,clip=true]
{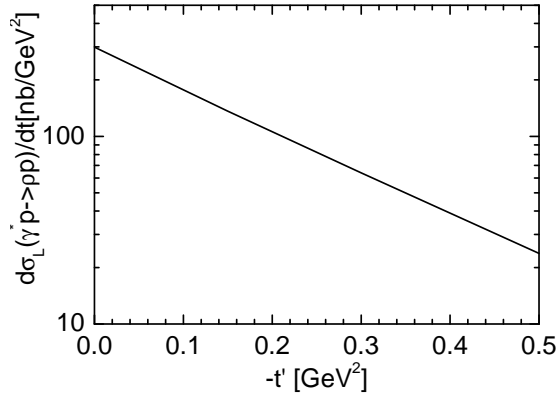}
\end{center}
\caption{The differential cross section for $\rho$ production versus
  $t'$ at $W=5\,\gev$ and $Q^2=4\,\gev^2$.} 
\label{fig:dsdt}
\end{figure}

 
In Fig.\ \ref{fig:sigmaL-lowW} we show the results  for $\sigma_L$ at
$W=5$ and $10\,\gev$ and compare them to the data from HERMES
\ci{HERMES-prel,hermes-rho} and the FERMILAB experiment E665 \ci{e665}. 
Again we observe good agreement with experiment. The slopes of the 
differential cross section are somewhat smaller at lower energies than 
those at the HERA energy shown in Fig.\ \ref{fig:slope}. For instance 
at $W=5\,\gev$ and $Q^2=4\,\gev^2$ we obtain $5.0\,\gev^{-2}$ for 
$\rho$ production and $4.8\,\gev^{-2}$ for the case of the $\phi$. As
yet the HERMES collaboration has only provided preliminary results for 
these slopes: $6.32\pm 0.72\,\gev^{-2}$ at $Q^2=3.7\,\gev^2$ for 
$\rho$ \ci{tytgat} and $4.6\pm 1.2\,\gev^{-2}$ averaged over the range 
$0.7 \le Q^2 \le 5\,\gev^2$ for $\phi$ production~\ci{HERMES-slopes-prel}. 
A slope for $\rho$ production that is considerably larger than that 
for $\phi$ production is difficult to get in the handbag approach. 
Although it seems tempting to assign such an effect to the valence
quark contribution by choosing a non-zero value for $b_{\rm val}$ this 
is likely not the solution since it would lead to a Dirac form factor 
that drops too fast with $t$. An alternative possibility seems to
change the $t'$ dependence of the sea quark GPD by choosing a value for 
$b_{\rm sea}$ larger than that for $b_g$ which is theoretically not
forbidden. On account of the different weights of the sea contribution 
in both the processes of interest this may generate a somewhat larger 
slope in the case of the $\rho$. In regard to the present experimental
situation we leave this question unanswered for the time being. 
Pertinent future data may settle this issue. 
In order to demonstrate how close the $t'$ dependence of the
differential cross section to an exponential behavior is we show 
in Fig.\ \ref{fig:dsdt} the cross section for $\rho$ production at
$W=5\,\gev$ and $Q^2=4\,\gev^2$ as an example.  

\begin{figure}[t]
\begin{center}
\includegraphics[width=.42\textwidth,bb= 28 310 528 700,clip=true]
{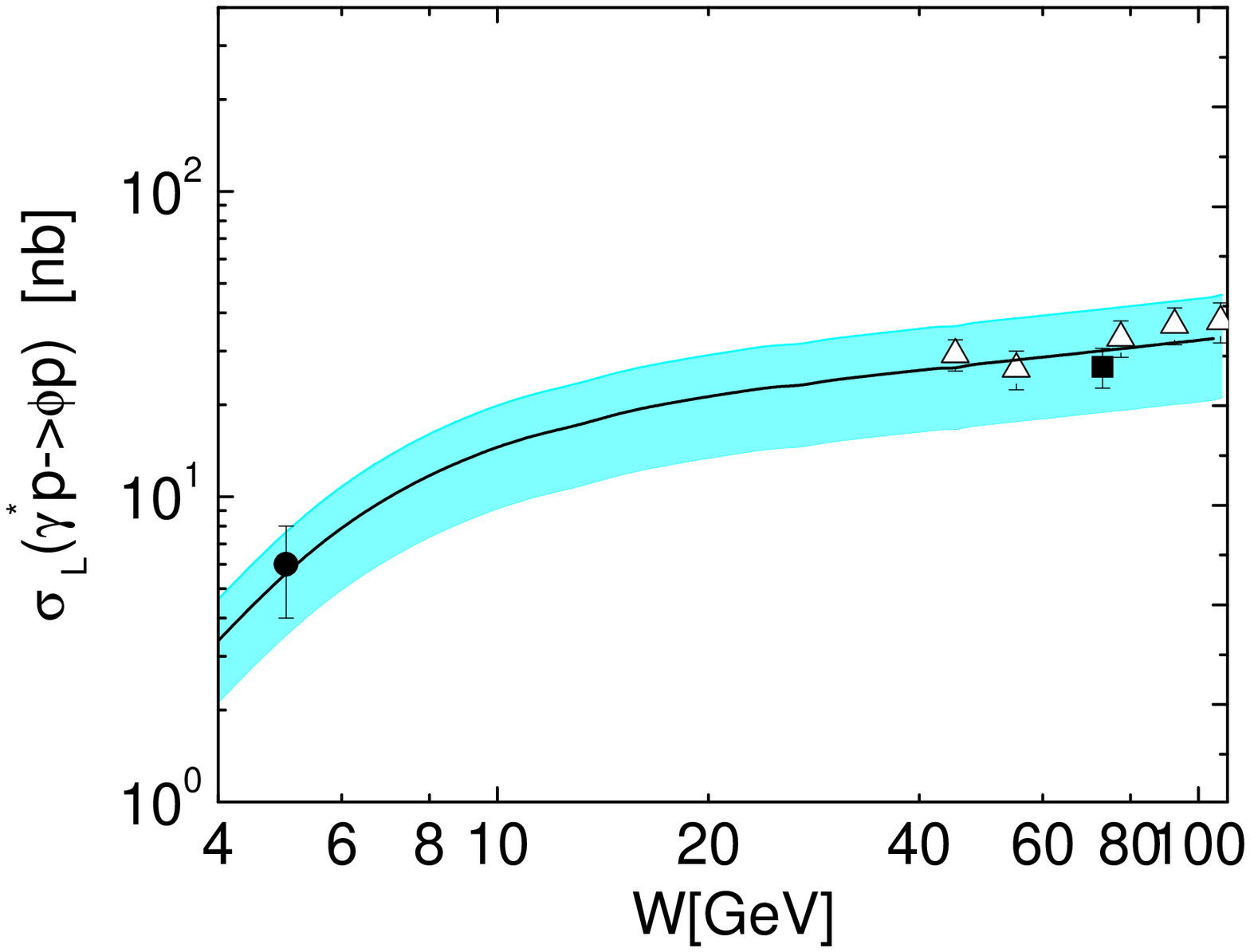}\hspace*{1.0em}
\includegraphics[width=.41\textwidth, bb=38 311 528 698,clip=true]
{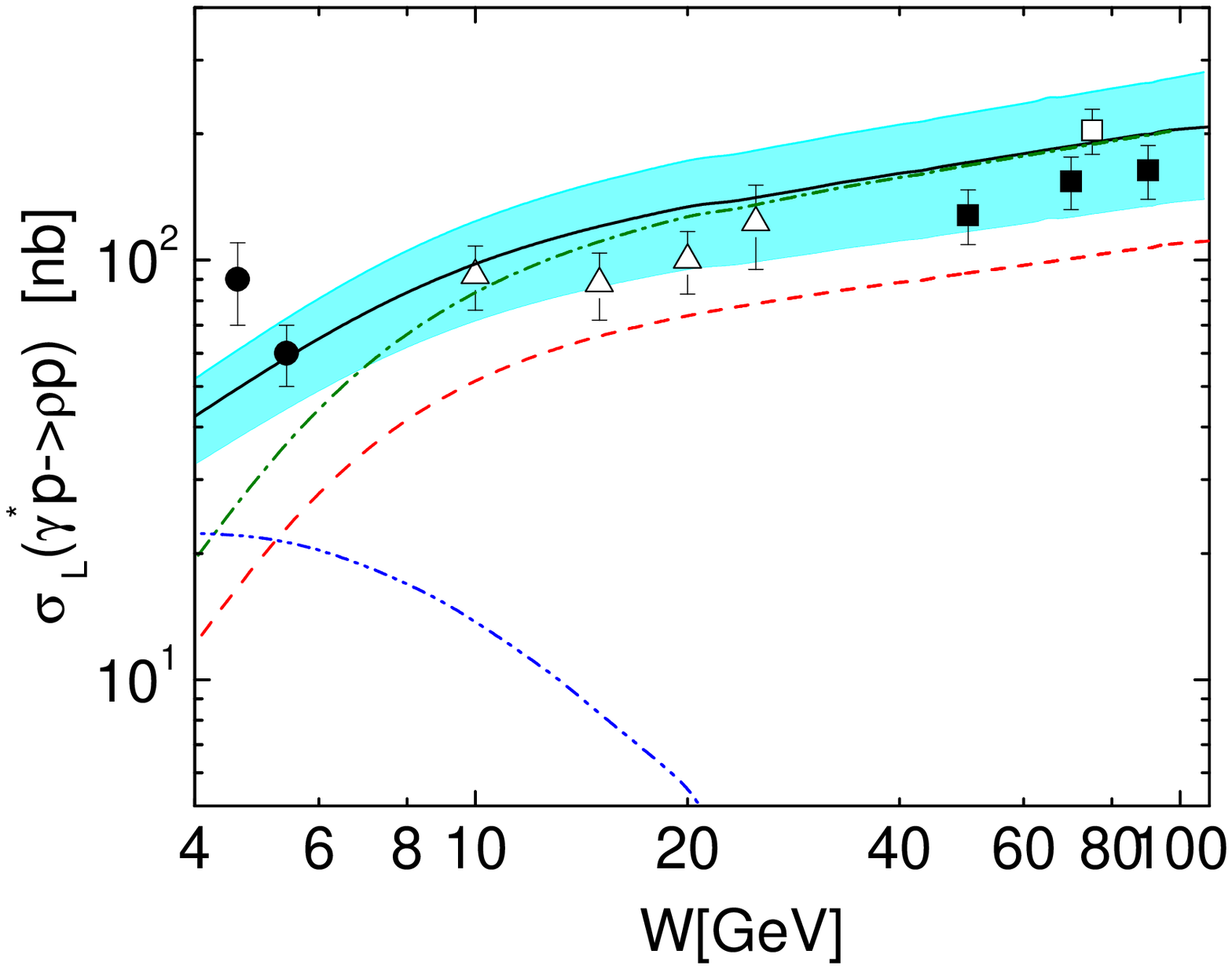}
\end{center}
\caption{The longitudinal cross section for $\phi$ (left) and $\rho$  
(right) electroproduction versus $W$ at $Q^2=3.8\,\gev^2$ and 
$4\,\gev^2$, respectively. The handbag predictions are evaluated
from the interval $-t'\leq 0.5\,\gev^2$. Data for $\phi$ production
are taken from HERMES \ci{HERMES-prel} (solid circle), ZEUS \ci{zeus05} 
(open triangles) and H1 \ci{adloff} (solid square). The data for $\rho$
production are taken from HERMES \ci{hermes-rho} (solid circles), E665
\ci{e665} (open triangles), ZEUS \ci{zeus98} (open square) and H1
\ci{h1} (solid square). The dashed (dash-dotted, dash-dot-dotted) line
represents the gluon (gluon + sea, (gluon + sea)-valence interference plus valence
quark) contribution. For other notations cf.\  Fig.\ \ref{fig:sigmaL-Q}.}
\label{fig:sigmaL-W}
\begin{center}
\includegraphics[width=.47\textwidth,bb= 10 295 581 700,clip=true]
{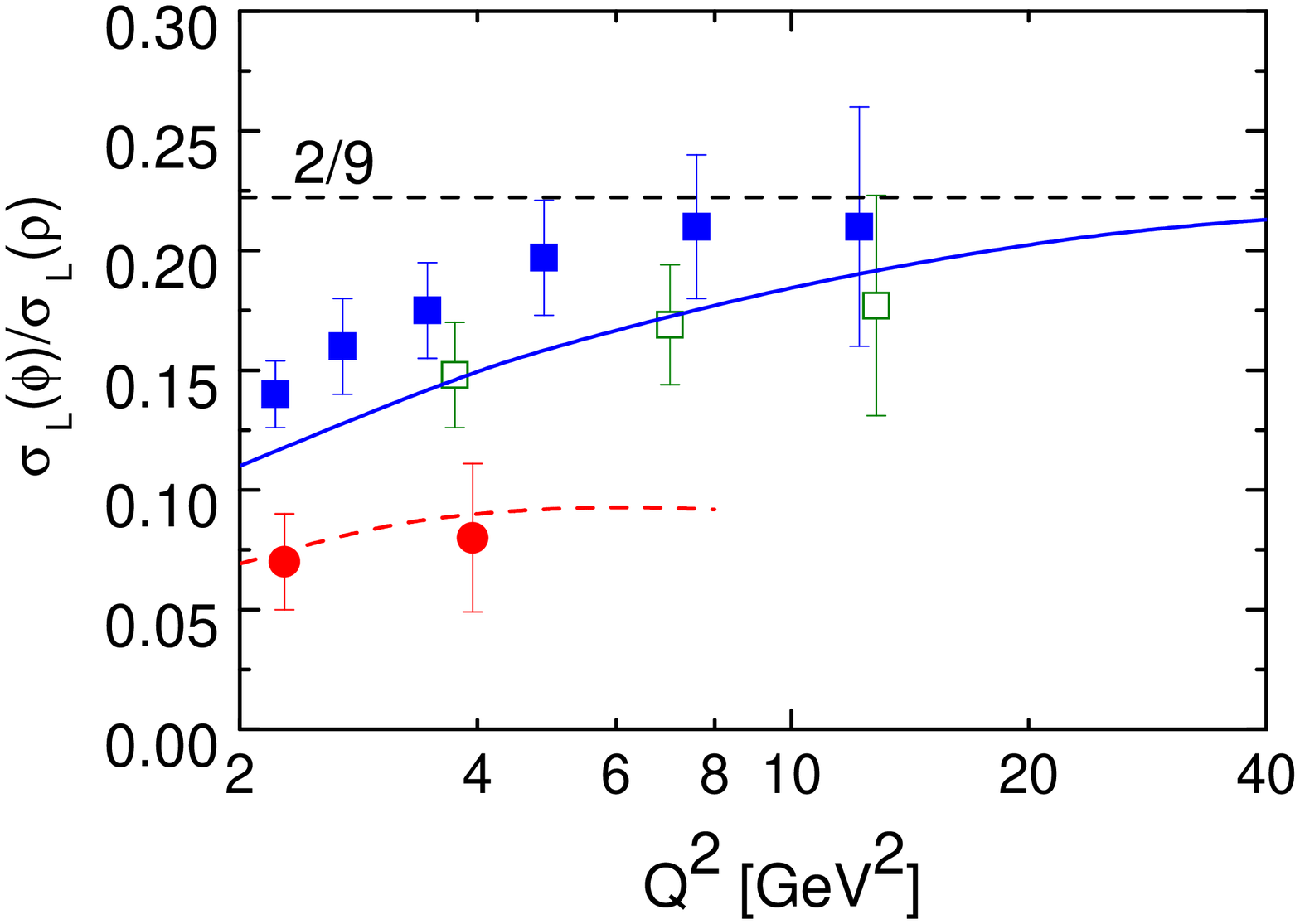}\hspace*{1.0em}
\includegraphics[width=.395\textwidth, bb=38 290 526 710,clip=true]
{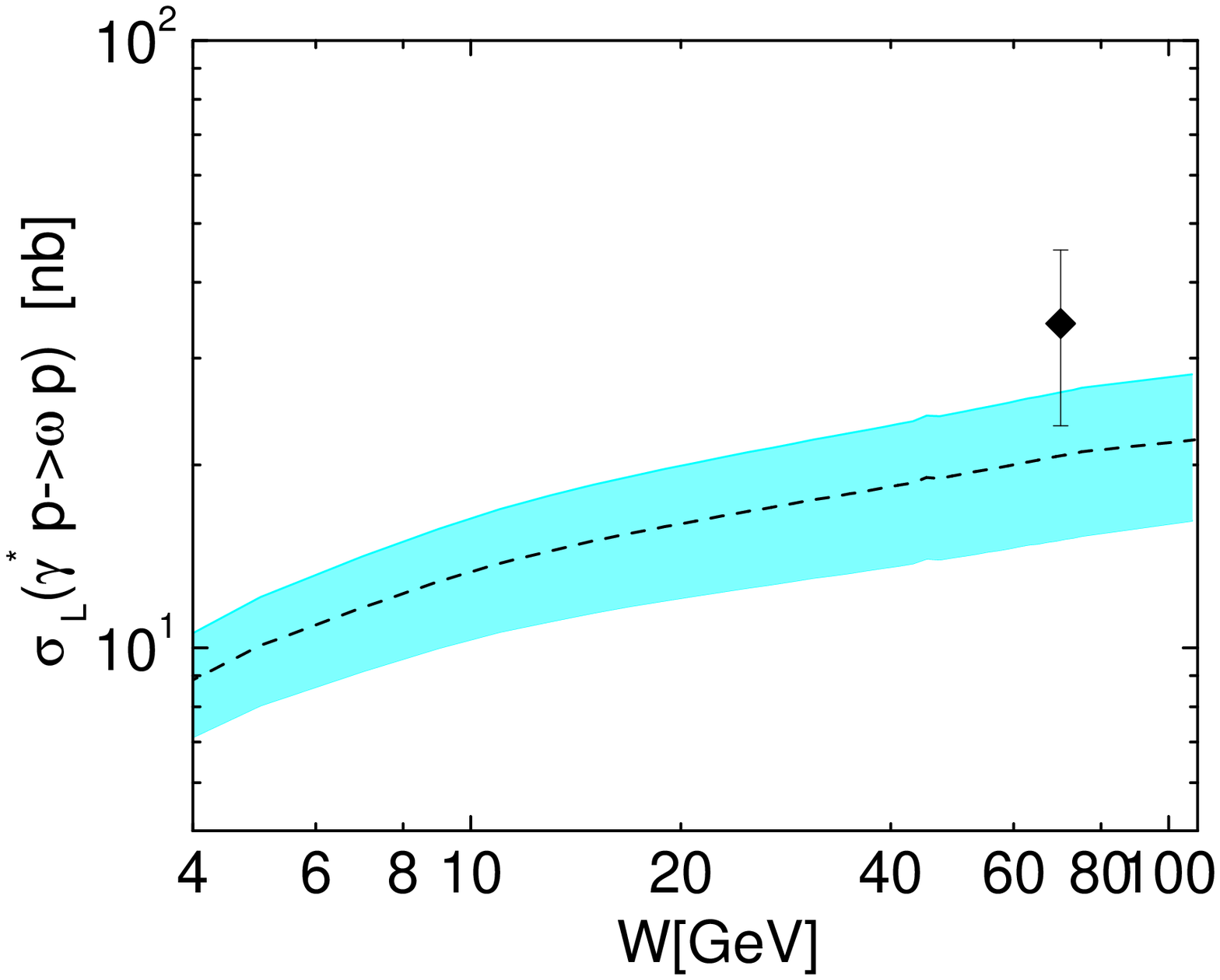}
\end{center}
\caption{Left: The ratio of the longitudinal cross sections for $\phi$
and $\rho$ production. Data are taken from H1 \ci{h1,adloff} (solid squares), 
ZEUS \ci{zeus98,zeus05} (open squares) and HERMES
\ci{HERMES-prel,hermes-rho} (solid circles). The solid (dashed) line
represents the handbag predictions at $W=75 (5)\, \gev$. Right:
Predictions for $\omega$ electroproduction versus $W$ at $Q^2=3.5\,\gev^2$.
For comparison the full cross section for $\omega$ production,
measured by ZEUS \ci{zeus-omega}, is also shown (solid diamond).}
\label{fig:omega}
\end{figure}

Consulting Fig.\ \ref{fig:sigmaL-W} one sees that also the $W$ dependence
of the longitudinal cross section is correctly described within the
handbag approach. Note that $\sigma_L$ in Figs.\ \ref{fig:sigmaL-Q},
\ref{fig:sigmaL-Q-rho}, \ref{fig:sigmaL-lowW} and \ref{fig:omega} (left)
is obtained by 
integrating the differential cross section over the range of $t'$ 
that is used in the various experiments. Thus, we integrate from 0 
to $-0.6, -0.5, -0.4\, \gev^2$ for HERA, COMPASS (and E665), and 
HERMES kinematics, respectively. For $\sigma_L$ in Figs.\
\ref{fig:sigmaL-W} and \ref{fig:omega} (right) on the 
other hand we integrate up to $-0.5\,\gev^2$ throughout but compare 
with actual data. The cross sections exhibit kinks at about 
$W=10\,\gev$, rather markedly for $\phi$ production, milder in the 
case of the $\rho$. These kinks are related to the sharp fall off 
of the gluon and sea quark GPDs with increasing $\xi$ (note that 
$\xi \propto 1/W^2$ at fixed $Q^2$), see the gluon PDF shown in 
Fig.\ \ref{fig:gluon}. The additional valence quark contribution in 
$\rho$ production 

\clearpage
\noindent
mitigates the kink. For $\rho$ production we also 
show in Fig.\ \ref{fig:sigmaL-W} the individual contributions from the 
gluons, sea and valence quarks. At $W\simeq 5\,\gev$ the latter 
contribution (including the interference with the gluon and sea
quarks) amounts to about $50\%$ of the full result but decreases 
rapidly with increasing $W$. It only contributes about $10\%$ at 
$W=10\,\gev$ and is negligible at the HERA energy.  

All the results we presented so far are evaluated from the Gaussian
\wf{} \req{wavefunction}. One may wonder what the consequences of other
choices of the \wf{} are. In order to provide a partial answer to this
question we multiply the \wf{} \req{wavefunction} for the $\rho$ meson
with the factor
\be
1 + B_2(\mu_0)\,\left[\frac{\als(\mu_F)}{\als(\mu_0)}\right]^{50/81}\, 
       C_2^{3/2}(2\tau-1)\,.
\label{eq:Gegenbauer}
\ee 
It corresponds to the first two terms of the expansion of the
meson \da{} upon the Gegenbauer polynomials $C_n^{3/2}$, the
eigenfunctions of the evolution kernel for mesons \ci{brodsky}. For an
estimate of the impact of that factor on the cross section for $\rho$
production we adopt the value
\be
B_2(\mu_0=1\,\gev) \= 0.18\,,
\label{b2-sumrules}
\ee
obtained from QCD sum rules for the expansion coefficient \ci{braun}. 
From this value of $B_2$ we find that the cross section for $\rho$ 
production increases by approximately $14\, (24)\%$ at $Q^2=4\,
(20)\,\gev^2$ in the entire range of energy we examine. This is well 
within the uncertainties of our approach which are represented by the 
error bands shown in the various figures. The gradual decrease of the 
second Gegenbauer term with increasing scale is to a large extent 
compensated by the diminishing suppression through the quark
transverse momenta and the Sudakov factor. In a calculation to 
leading-twist order the effect of the higher Gegenbauer terms is more 
pronounced at low $Q^2$. For instance, at $Q^2=4\,\gev^2$ the $\rho$ 
cross section increases by $31\%$ to leading-twist order using
\req{b2-sumrules} and $\mu_F=Q$.

In the limit of negligible valence quark contributions one has the
following relative strength of the three cross sections:
$\rho\, : \,\omega\, : \, \phi = 1\, :\, 9\, :\, 2/9$
up to flavor symmetry breaking effects as the differences in the decay
constants and in the \wf{} or distribution amplitudes or the flavor 
symmetry breaking factor \req{eq:kappas} in the sea GPDs. While the
latter two effects disappear for $Q^2\to \infty$ due to evolution,
the differences in the decay constants are scale independent. Hence,
for very large scales the handbag approach predicts
\be
\sigma_L(\phi)/\sigma_L(\rho) \Longrightarrow 
           \frac{2}{9}\,\left(\frac{f_\phi}{f_\rho}\right)^2 \= 0.248\,.
\label{eq:ratio}
\ee
An analogous result holds for the ratio of the $\omega$ and $\rho$
cross sections. The HERA data \ci{h1,zeus98,zeus05,adloff} shown in
Fig.\ \ref{fig:omega}, are not far from the symmetry limit $2/9$ especially 
at larger values of $Q^2$ but the face values are clearly below it. 
The evolution effect of a $\rho$ wave function broader in $\tau$ than 
the Gaussian given in Eq.\ \req{wavefunction} is too mild if one
accepts the QCD sum rule estimate of the Gegenbauer coefficients, see 
above discussion. It does not explain the $Q^2$ dependence of the
data. Thus, we are compelled to conclude that the bulk of 
the effect seen in the HERA data is due to flavor symmetry breaking in 
the sea. Indeed with $\kappa_s$ as given in Eq.\ \req{eq:kappas} we 
obtain the results for the ratio of the $\phi$ and $\rho$ cross
sections shown in Fig.\ \ref{fig:omega} which agree fairly well with 
experiment. The cross section ratio at the HERMES energy is smaller as
at the higher HERA energy which as a glance at Eq.\ \req{eq:lt-amp} 
reveals, is to be assigned to the additional valence quark
contribution to the $\rho$ cross section at $W=5\,\gev$ (see also 
Ref.\ \ci{vinnikov}). Indeed our results are in agreement with the 
HERMES data \ci{HERMES-prel,hermes-rho}, see Fig.\ \ref{fig:omega}. 
    
Predictions for $\omega$ production are presented in Fig.\
\ref{fig:omega} as well. They are obtained by assuming $a_\omega=a_\rho$.
Only the unseparated cross section has been measured by the ZEUS 
collaboration \ci{zeus-omega} at $W=70\,\gev$ and $Q^2=3.5$ and 
therefore we cannot directly compare with our results. However,
assuming that the ratio of the longitudinal and transversal cross
sections is about 2 at this kinematics, one expects a longitudinal 
cross section that amounts to about 2/3 of the full cross section and
this would be in agreement with our result. The valence quark
contribution is stronger for $\omega$ than for $\rho$ production as is
expected from Eq.\ \req{eq:lt-amp}. For our kinematical point of
reference, $W=5\,\gev$ and $Q^2=4\,\gev^2$, for instance, it amounts
to about $65\%$ of the $\omega$ cross section. Data on the longitudinal
cross section for $\omega$ production would be highly welcome. They
would provide information on a second combination of the $u$ and $d$ quark
GPDs $e_u\,H^u+ e_d\,H^d$.
\section{Summary}
\label{sec:conclusions}
We have investigated light vector-meson electroproduction within the
handbag approach. The partonic subprocesses are treated within the
modified perturbative approach and the GPD $H$ for gluons, sea and 
valence quarks is constructed from the CTEQ6M PDFs through double 
distributions using a Regge-inspired $t$ dependence. The GPDs $H$
respect all theoretical constraints, i.e.\ the reduction formulas, 
positivity and polynomiality as well as the sum rule for the Dirac form
factor of the proton. From our approach we have obtained a fair 
understanding of the longitudinal cross section over a large range of 
energy (from $W=5$ till $100\,\gev$) and photon virtualities ( from 
2.5 to $40\,\gev^2$) provided  $\xbj$ is small. A remarkable outcome
of our investigation is that the gluon contribution play an important
role over the entire range of energy we have explored. The sea quarks
also contribute considerably at all energies although to a lesser
extent than the gluons. The valence quarks on the other hand are only
of importance for $\rho$ and $\omega$ production and for energies less 
than about $10\,\gev$.  

We have simplified the parameterization of the sea quark
GPDs, not all details of the PDFs are transfered to them. Thus, 
we have reduced the sea quark GPDs to a
single function allowing for differences only through a flavor
symmetry factor. At the present stage of our knowledge on the GPDs a
more refined model would provide a pseudo accuracy that does not meet
the uncertainties of the GPD model and that would rather confuse than
elucidate. With our sea quark GPDs it becomes evident that the $Q^2$
dependence of the $\phi$ - $\rho$ ratio of the longitudinal cross
sections at HERA energies is generated by the flavor symmetry breaking
in the sea. Data for $\omega$ production may  perhaps force us to
improve the parameterization of the sea quark GPDs. The data for
$\rho$ production alone only probe the combination $e_u H^u_{\rm sea}
-e_d H^d_{\rm sea}$. 

In our previous work \ci{first} we have also calculated the amplitudes for 
other transitions from the virtual photon to the vector meson within
the handbag approach. With them we have achieved a fair description of the 
transverse cross section and the spin density matrix elements at HERA 
energies where the gluon contribution dominates. The application of 
this approach to the spin density matrix elements at lower energies 
necessitates the inclusion of the quark contribution into the
analysis which is left to a forthcoming paper.     

The handbag approach with the reggeized GPDs bears similarities to
other theoretical models for vector-meson electroproduction which are
mainly applied to the high energy and/or very low $\xbj$ regime. The
gluonic subprocess, see Fig.\ \ref{fig:feynman}, with the accompanied
proton-gluon vertex function forms the basis of many models. The
various approaches essentially differ in the treatment of that vertex
function (Pomeron residue \ci{sandy,donn} or the gluon PDF in the BFKL
color dipole model \ci{ivanov04} and in the leading-log approximation
\ci{fra95,mrt}) 
and in the assumptions which of the partons in the Feynman graphs are 
considered as soft, i.e.\ as quasi on-shell, and which as hard. In the 
handbag approach with the QCD factorization theorems \ci{rad96,col96} 
as foundation, the protons emit and reabsorb quasi on-shell partons
and the associated vertex function, a soft proton matrix element, is 
regarded as a GPD. The quark and antiquark entering the final state 
meson  are also considered as on-shell particles. The associated soft 
$q\bar{q}\to V$ transition is parameterized as a light-cone \wf{} or 
distribution amplitude. All other partons in the graphs shown
in Fig.\ \ref{fig:feynman} are highly virtual. The advantage of the 
handbag approach is that once the GPDs are fixed other hard exclusive 
processes, as for instance deeply virtual Compton scattering, can be 
predicted.

\section*{Acknowledgements} 
We thank J.\ Pumplin for comments on the PDFs, A.\ Borissov for
discussions and the HERMES collaboration for permission to use 
preliminary data. This work has been supported in part by the 
Russian Foundation for Basic Research, Grant 06-02-16215, the 
Integrated Infrastructure Initiative ``Hadron Physics'' of the 
European Union, contract No. 506078 and by the Heisenberg-Landau 
program. 

\end{document}